\documentclass[preprint, twocolumn]{aastex631}
\usepackage{natbib}[1999/02/25]
\usepackage{rotating}
\usepackage{booktabs}

\usepackage{SIunits}

\usepackage{paralist}

\usepackage{psfrag,color}
\shorttitle{Kinematic Properties of CVs}
\shortauthors{Canbay et al.}
\graphicspath{{./}{Figures/}}

\begin{document}
	\title{Kinematics of Cataclysmic Variables in the Solar Neighborhood in the {\it Gaia} Era}

\correspondingauthor{Remziye Canbay}
\email{rmzycny@gmail.com}

\author[00000-0003-2575-9892]{Remziye Canbay}
\affiliation{Istanbul University, Institute of Graduate Studies in Science, Programme of Astronomy and Space Sciences, 34116, Istanbul, Turkey}

\author[0000-0002-0688-1983]{Tansel AK}
\affiliation{Istanbul University, Faculty of Science, Department of Astronomy and Space Sciences, 34119, Istanbul, Turkey}
\affiliation{Istanbul University, Observatory Research and Application Center, Istanbul University, 34119, Istanbul, Turkey}

\author[0000-0003-3510-1509]{Sel\c{c}uk Bilir}
\affiliation{Istanbul University, Faculty of Science, Department of Astronomy and Space Sciences, 34119, Istanbul, Turkey}

\author[0000-0002-5141-7645]{Faruk Soydugan}
\affiliation{Faculty of  Sciences, Department of Physics, Canakkale Onsekiz Mart University, Canakkale, Turkey}
\affiliation{Astrophysics Research Center and Ulupınar Observatory, Canakkale Onsekiz Mart University, Canakkale, Turkey}

\author[0000-0003-1883-6255]{Zeki Eker}
\affiliation{Akdeniz University, Faculty of Sciences, Department of Space Sciences and Technologies, 07058, Antalya, Turkey}

	
	

\begin{abstract}
\noindent
Using high-precision astrometric data from {\it Gaia} DR3 and updated systemic velocities from the literature, kinematical properties of cataclysmic variables (CVs) were investigated. By constraining the data according to the total space velocity error and Galactic population class, a reliable sample of data was obtained. Non-magnetic CVs located in the thin disk have been found to have a total space velocity dispersion of $\sigma_{\nu} = 46.33\pm4.23$ km s$^{-1}$, indicating that the thin disk CVs with a mean kinematical age of $\tau = 3.95\pm0.75$ Gyr are much younger than the local thin disk of the Galaxy with $\tau\sim$6-9 Gyr. Total space velocity dispersions of non-magnetic CVs belonging to the thin disk component of the Galaxy were found to be $\sigma_{\nu}=47.67\pm3.94$ and $\sigma_{\nu}=44.43\pm4.33$ km~s$^{-1}$ for the systems below and above the orbital period gap, respectively, corresponding to kinematical ages of $\tau=4.19\pm0.71$ and $\tau=3.61\pm0.74$ Gyr. $\gamma$ velocity dispersions of the thin disk CVs below and above the gap were obtained $\sigma_{\gamma} = 27.52\pm2.28$ and $\sigma_{\gamma} = 25.65\pm2.44$ km s$^{-1}$, respectively. This study also shows that the orbital period is decreasing with increasing age, as expected from the standard theory. The age-orbital period relation for non-magnetic thin disk CVs was obtained as $dP/dt=-2.09(\pm0.22)\times10^{-5}$ sec yr$^{-1}$. However, a significant difference could not be found between the $\gamma$ velocity dispersions of the systems below and above the gap, which were calculated to be $\sigma_{\gamma} = 27.52\pm2.28$ and $\sigma_{\gamma} = 25.65\pm2.44$ km s$^{-1}$, respectively.
\end{abstract}

\keywords{Star: Cataclysmic binaries, Galaxy: Stellar dynamics and kinematics: Galaxy: Solar neighborhood}

\section{Introduction}

Cataclysmic variables (CVs) are a class of binaries consisting of a white dwarf primary and a secondary, typically a late-type main sequence star. The secondary fills its Roche lobe and transfers material to the white dwarf through an accretion disk, leading to a variety of phenomena such as novae, dwarf novae, and nova-like variables with irregular changes in brightness due to the thermal instability of the accretion disk \citep{Warner1995, Hellier2001, Takata2022}. CVs are a major focus of interest in the field of stellar astrophysics because of their complex evolutionary pathways, their variability, and the information they provide about binary star evolution and accretion processes  \citep{Warner1995,Hellier2001,Giovannelli2008,Knigge2011a}.

The fundamental nature of CV systems is intrinsically linked to their binary configuration. Mass transfer from the donor star to the white dwarf is driven by the angular momentum loss reducing the distance between the primary and the secondary and causing the Roche lobe overflow of the secondary star \citep{Knigge2011a}. Magnetic braking and gravitational radiation could be the two mechanisms to maintain such an angular momentum loss associated with the decreasing of the system size. It has been well established that gravitational wave radiation leads to angular momentum loss in CVs at short orbital periods, and some form of magnetic braking at longer periods \citep{Rappaportetal1983,Schreiberetal2016}. Then, the mass transfer, which could be sporadic or continuous, leads to the rich phenomenology associated with CVs, including outbursts and periodic variability \citep{Patterson1984}. 

The number of CVs appears to decrease in the orbital period range of $2<P_{\rm orb}({\rm h})<3$, which is known as the period gap. The standard evolutionary model of CVs successfully explains why such a period gap exists. In addition, population synthesis studies based on the standard model predict an accumulation of systems below the period gap, as the CVs spend most of their lifetime below the gap \citep{Kolb1999, King2002, Littlefair2008}. Indeed, SDSS revealed many systems near the period minimum \citep{Gansickeetal2009}, though only a handful of systems past the period minimum, known as period bouncers, have been seen \citep{Inightetal2023}. However, there is a disagreement between observed and predicted minimum orbital period values, 76-82 \citep[see][and references therein]{Kalomeni2016} and 65-70 minutes \citep{Knigge2011b,  McAllister2019}, respectively. Another problem is about white dwarf masses. Single white dwarf masses are smaller than those of white dwarfs in CVs \citep[see][and references therein]{Zorotovic2020}, while the white dwarf mass in CVs does not depend on the orbital period \citep{McAllister2019,Pala2022}. In addition, the observed space density of these systems is in agreement only with the lower limit of the predictions of the standard model \citep[][and references therein]{Canbay2023,Rodriguezetal2024}. In a recent study, \citet{Schaefer2024} investigated the orbital period changes of CVs and found that nearly half of the objects have increasing orbital periods. This is impossible in the magnetic braking model which is used as the angular momentum loss mechanism for the objects above the orbital period gap in the standard evolutionary model of CVs. Thus, \citet{Schaefer2024} claimed that the magnetic braking model fails in explaining the orbital period changes of these systems. \cite{Kingetal2024}, however, proposed a counterargument that short-term phenomena can result in the observed period changes. Additional angular momentum loss mechanisms were  also introduced  \citep{Andronov2003, Taam2001, Schenker2002, Willems2005, Willems2007,Knigge2011a,AlBandryetal2022}. Note that evolutionary scenarios of magnetic CVs may be different from scenarios of non-magnetic CVs \citep[see the references in][]{Belloni2020a}. 

CV evolution models are often tested through photometric data sets, though these are typically skewed due to selection biases, particularly those influenced by brightness \citep{Pretorius2007}. Despite this, the age distribution of CVs remains unaffected by brightness-related selection biases \citep{Kolb2001}, as the age of a CV does not impact its mass transfer rate for a given orbital period. 
 
The modelled population of Galactic CVs based on the standard formation and evolution model predicted that systems above the period gap have ages younger than 1.5 Gyr, with an average age of 1 Gyr for them, while the mean age of the ones below the period gap is 3-4 Gyr, with all ages above about 1 Gyr \citep{Kolb1996, Ritter1986}. \citet{Kolb1996} predicted dispersions of the $\gamma$ velocities to be $\sigma(\gamma)\simeq 15$  and $\sigma(\gamma)\simeq 30$ km s$^{-1}$ for the systems above and below the period gap, respectively, where $\gamma$ is the centre of mass radial velocity of a CV. They used the empirical relation between age $t$ and space velocity dispersion $\sigma(t)$ of field stars in the solar neighborhood \citep{Wielen1977, Wielen1992}, $\sigma(t)$= $k{\rm_1}$ + $k{\rm_2}$ $t^{1/2}$ ($k{\rm_1}$ and $k{\rm_2}$ are constants), to estimate the dispersions of $\gamma$ velocities. According to \citet{Kolb2001}, the age difference of systems above and below the orbital period gap should be mainly due to time spent evolving from the post-common envelope (post-CE) orbit into contact. Kolb's (\citeyear{Kolb2001}) another prediction is that the $\gamma$ velocity dispersions of CVs are $\sigma(\gamma)\simeq 27$  and $\sigma(\gamma)\simeq 32$ km s$^{-1}$  for the systems above and below the gap, respectively, if magnetic braking does not operate in the detached phase during the evolution. 

However, observed $\gamma$ velocities for the systems collected from published radial velocity studies do not show such a difference between velocity dispersions of the CVs above and below the period gap \citep{Van1996}. Precise $\gamma$ velocity measurements for only four dwarf novae above the period gap performed by \citet{North2002} suggest a velocity dispersion of $\sim$8 km s$^{-1}$. \citet{Ak2010} estimated kinematical ages of the non-magnetic CVs as 5.01$\pm$1.48 and 3.65$\pm$1.34 Gyr for the systems below and above the gap. Although their difference is in agreement with that predicted by \citet{Kolb1996}, these ages are older than expected. \citet{Ak2010} also calculated $\gamma$ velocity dispersions of $\sigma_{\gamma}$ = 26$\pm$4 and $\sigma_{\gamma}$ = 30$\pm$5 km s$^{-1}$ for non-magnetic systems above and below the gap, respectively. The observational difference between these $\gamma$ velocity dispersions is smaller than that expected from the modelled population study of CVs made by \citet{Kolb1996}. It is interestingly in agreement with the above prediction of \citet{Kolb2001}. \citet{Ak2015} increased the number of systems with $\gamma$ velocities by collecting data from the literature and performing a new kinematic analysis. Their difference from \citet{Ak2010} was not only the number of systems used but also estimating kinematic parameters for CVs belonging to the thin disk component of the Galaxy, which constitute a more homogeneous group. \citet{Ak2015} estimated mean kinematical ages of the non-magnetic thin disk CVs as 3.40$\pm$1.03 and 3.90$\pm$1.28 Gyr for the systems below and above the period gap. This estimation is not in agreement with that found by \citet{Ak2010} for the systems below the gap. \citet{Ak2015} calculated $\gamma$ velocity dispersions of $\sigma_{\gamma}$ = 24.95$\pm$3.46 and $\sigma_{\gamma}$ = 26.60$\pm$4.48 km s$^{-1}$ for non-magnetic thin disk systems below and above the gap, respectively, which are very similar to each other. They also found $\gamma$ velocity dispersions of $\sigma_{\gamma}$ = 32.10$\pm$4.41 and $\sigma_{\gamma}$ = 25.92$\pm$4.03 km s$^{-1}$ for magnetic and non-magnetic thin disk CVs, respectively, corresponding mean kinematical ages of 5.64$\pm$1.39 and 3.69$\pm$1.22 Gyr. 

The advent of the {\it Gaia} mission \citep{Gaia2016_mission} has revolutionized our understanding of stellar populations in the Milky Way. With its unprecedented astrometric precision, {\it Gaia} has provided detailed and accurate trigonometric parallax/proper motion measurements, and photometric data for millions of stars, including CVs. 
This wealth of data has opened new avenues for studying the kinematics of CVs, allowing for more accurate distance determinations and motion analyses within the solar neighborhood \citep{Gaia2018, Gaia2021, Gaia2023}. These data can be used to refine the space velocity dispersion of CVs and compare them with the kinematics of field stars and other stellar populations. The kinematic ages of CVs can be estimated by comparing their space velocity dispersions with those of other stellar populations with known ages. Older populations tend to have higher velocity dispersions due to the cumulative effects of gravitational interactions with other stars and molecular clouds \citep{Wielen1977}. The centre of mass radial velocity of a CV ($\gamma$ velocity) is the next important parameter to estimate its space velocity and it must be collected from the literature. 

In this paper, we calculate the space velocities and derive the $\gamma$ velocity dispersions of CV groups according to different orbital period regimes and Galactic populations to test the model predictions. To do this, we have collected $\gamma$ velocities of these objects from the literature and used precise equatorial coordinate, trigonometric parallax and proper motion data from {\it Gaia} data release 3 \citep[Gaia DR3,][]{Gaia2023}. 

\section{Data}
Calculating the spatial velocity components of stars requires the equatorial ($\alpha,\delta$) and Galactic ($l, b$) coordinates of the objects, the proper motion components ($\mu_{\alpha} \cos \delta, \mu_{\delta}$), trigonometric parallaxes ($\varpi$), and the radial velocities. The systems classified as CV with known ${\gamma}$ velocities have been compiled from the literature, while astrometric data were retrieved from the {\it Gaia} DR3 \citep{Gaia2023}. Orbital periods  were collected from \citet[][Edition 7.24]{Ritter2003}, \citet{Downes2001} and literature.


\subsection{Systemic Velocities}
The space velocity of a star is computed using its radial velocity with respect to the Sun, which is measured from Doppler shifts of spectral lines. The systemic velocities ($\gamma$) of binary stars are used and calculated from the radial velocities. To predict the systemic velocity of a CV, the function $V_r(\phi)$ = $\gamma$ + $K_{1,2}$ sin$\phi$ is preferred as their orbit is circular. Here, $\phi$ is the orbital phase, $\gamma$ is the centre of mass radial velocity of a CV, $K_{1,2}$ represent semi-amplitudes of radial velocity variation and 1 and 2 denote primary (white dwarf) and secondary (donor) components of the system, respectively. Systemic velocity $\gamma$ is generally found by a non-linear least-square fit of this function to the observed radial velocities. 

The first collection for systemic velocities of CVs from the literature was created by \citet{Van1996} for CVs known up until 1994. Their study was followed by \citet{Ak2010} who collected systemic velocities of CVs from the literature up to the year 2007 in a similar style. \citet{Ak2015} increased the number of systemic velocities of CVs and analysed their data. \citet{Ak2010,Ak2015} used distances estimated from PLCs (Period-Luminosity-Colours) relations of \citet{Ak2007} and \citet{Ozdonmez2015}, respectively. 

In this study, the systemic velocity data for 455 cataclysmic variables have been collected from the literature, as well as checking previous compilations of \citet{Van1996} and \citet{Ak2010,Ak2015}. We adopted very similar criteria defined by \citet{Ak2010} when collecting  $\gamma$ velocities. Thus, if there is more than one measurement of $\gamma$ velocity for a system, we inspected the studies and found the one with the radial velocity curve including more data points that are well spread over the orbital phase. The $\gamma$ velocities obtained during superoutbursts of SU UMa-type dwarf novae were not taken into account. We adopted the $\gamma$ velocity recommended by a researcher when more than one $\gamma$ velocity determination was presented in a study. 

It should be noted that motions in the accretion disk or the matter stream falling on the disk from the secondary component of the system probably affect the spectral lines. Since Doppler shifts are measured from these lines,  $\gamma$ velocities determined from them may not be reliable. Although radial velocities measured from absorption lines represent the system best, they are weak lines and thus can not be observable in all CVs \citep{North2002}, as they are most likely produced in the secondary star's atmosphere. Furthermore, there are cases \citep[e.g.][]{Schwopeetal2011} in which the donor star's spectral line traces the irradiated surface of the donor star, not the centre of mass. That affects the amplitude of the radial velocity curve, but not the $\gamma$ velocity.

\begin{figure}
  \centering
  \includegraphics[width=\columnwidth]{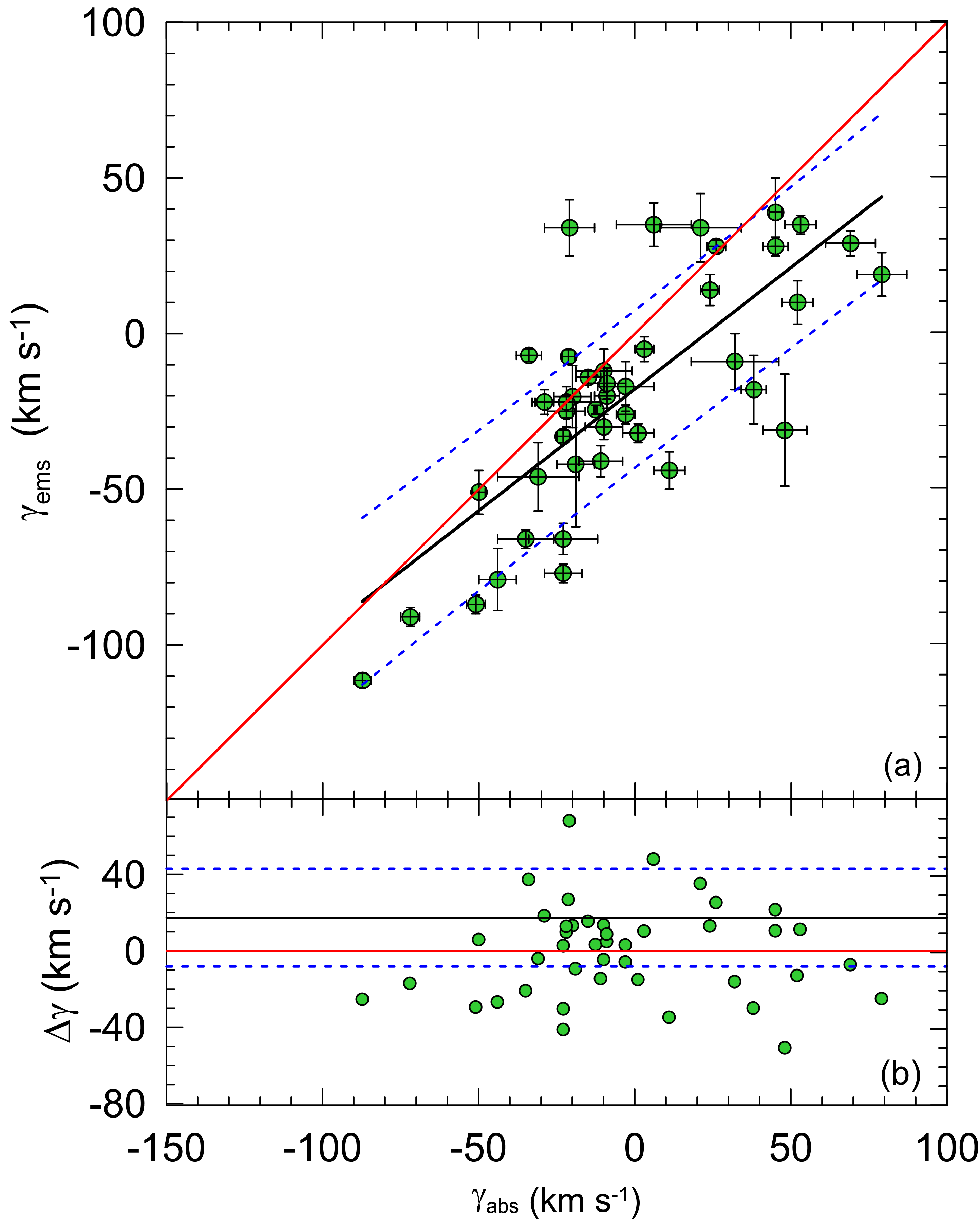}
   \caption{(a) Comparison of systemic velocities measured from emission ($\gamma_{\rm ems}$) and absorption ($\gamma_{\rm abs}$) lines for 44 CVs in the sample. Solid black line and blue dotted lines represent the linear fit and 1$\sigma$ limits, respectively. The red solid line in the upper panel is the one-to-one line. (b) Distribution of velocity differences. Red and black lines show zero level and mean value of velocity differences, respectively.}
   \label{fig:Ems-Abs}
\end{figure}

To understand the impact of using measurements from emission and absorption lines on the results, we compared $\gamma$ velocity measurements from emission ($\gamma_{\text{ems}}$) and absorption ($\gamma_{\text{abs}}$) lines. We found 44 CVs, for which $\gamma_{\text{ems}}$ and $\gamma_{\text{abs}}$ measurements are both available. These $\gamma$ values are compared in Figure \ref{fig:Ems-Abs}. As shown in Figure \ref{fig:Ems-Abs}, the systemic velocities calculated from emission and absorption lines are consistent within 1$\sigma$ errors, in general. The average difference of $\gamma_{\text{ems}}$ and $\gamma_{\text{abs}}$ values is $\langle \gamma_{\text{ems}}- \gamma_{\text{abs}} \rangle = +17.4\pm25.6$ km~s$^{-1}$, where the error is the standard deviation of individual differences. This average value is in agreement with those calculated by \citet{Van1996}, \citet{Ak2010} and \citet{Ak2015}, +2.5$\pm$13.8, +0.57$\pm$14.3 and -3.5$\pm$22.6 km~s$^{-1}$, respectively, within errors. We conclude that meaningful statistics can be obtained from the observed systemic velocities in our sample. Based on these comparisons, although $\gamma$ velocity measurements obtained from absorption lines were preferred during the compilation, we also collected $\gamma_{\text{ems}}$ velocities to construct our data sample. $\gamma$ velocities and associated uncertainties are listed in Table 1. Figure \ref{fig:inputerrors}a shows an error histogram of $\gamma$ velocities,  where intermediate polars (IPs) and polars (Ps) are shown separately. The median value and standard deviation of $\gamma$ velocity errors are 5.0 and 6.2 km~s$^{-1}$ for all CVs, respectively. 

\begin{figure}
  \centering
  \includegraphics[width=\columnwidth]{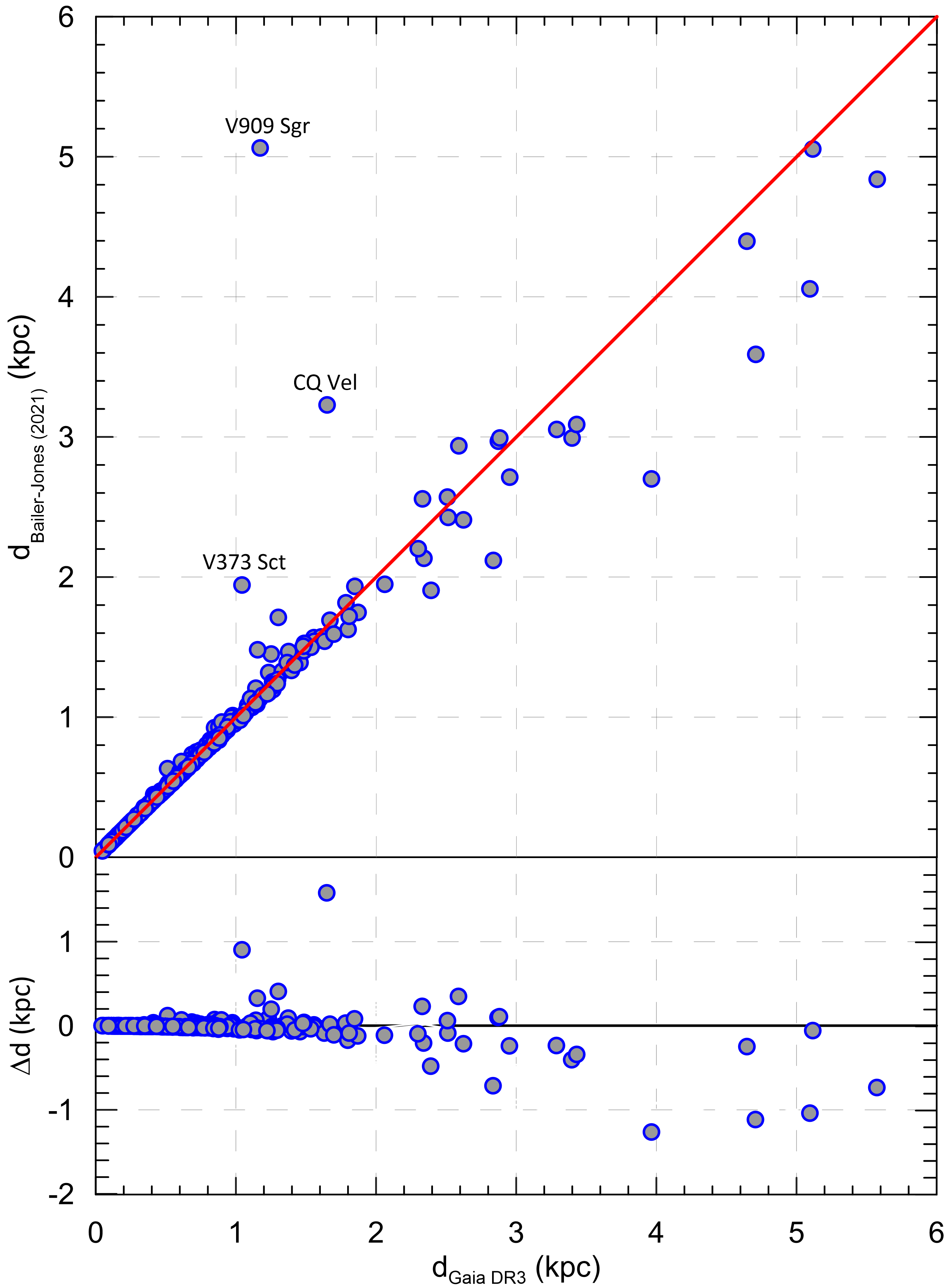}
   \caption{Comparison of distances calculated from {\it Gaia} DR3 \citep{Gaia2023} database and \citet{BailerJones2021} parallaxes for 432 CVs 
   in the sample. The distribution of differences is given in the lower panel.}
   \label{fig:distances}
\end{figure}

\begin{figure}
  \centering
  \includegraphics[width=\columnwidth]{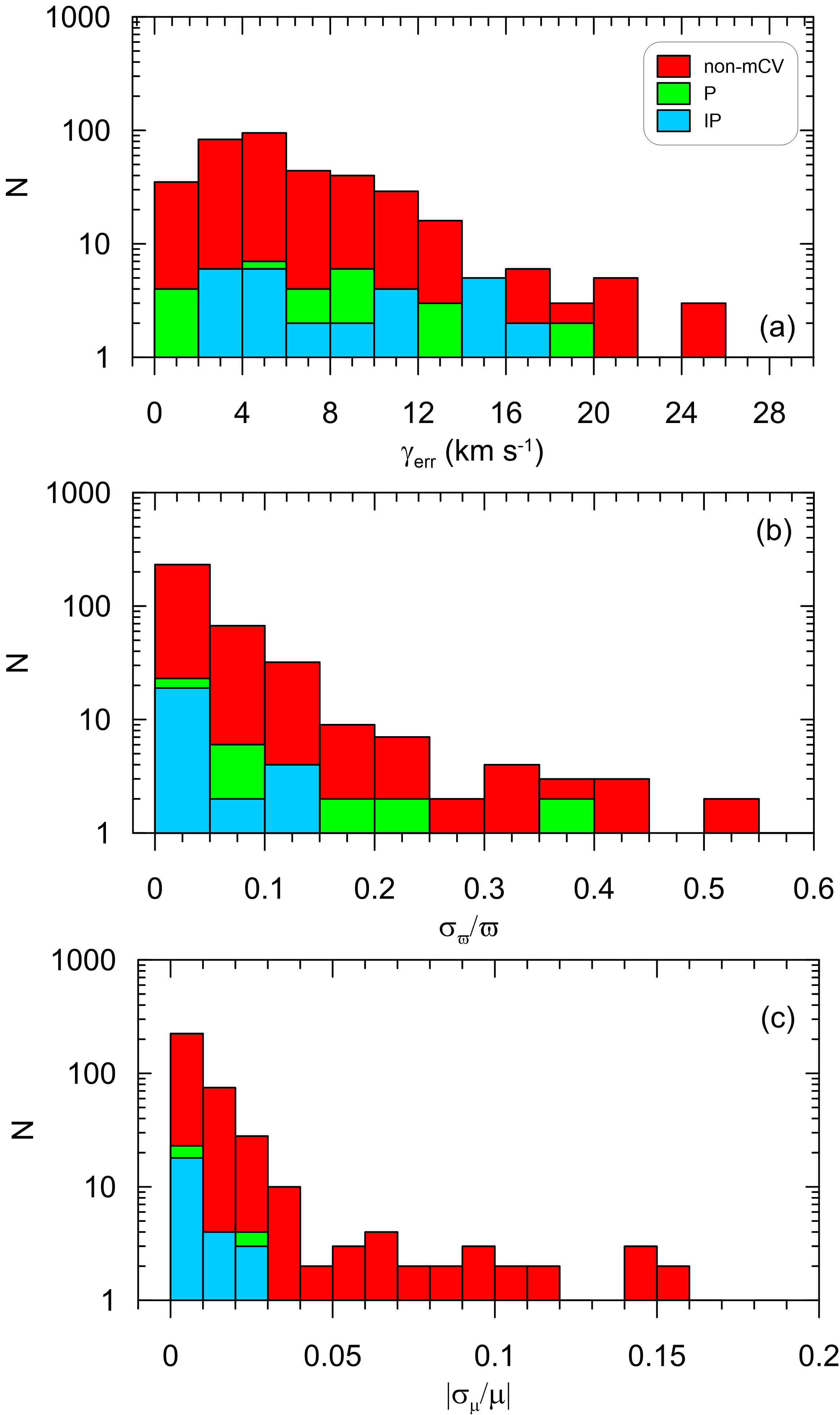}
   \caption{Distribution of $\gamma$ velocity errors (a), relative parallax errors (b) and relative proper motion errors (c) of the CV sample. P and IP denote polars and intermediate polars, respectively.}
   \label{fig:inputerrors}
\end{figure}

\subsection{Astrometric Data}

Equatorial coordinates ($\alpha,\delta$), proper-motion components ($\mu_{\alpha} \cos \delta, \mu_{\delta}$) and trigonometric parallaxes ($\varpi$) for 455 CVs with known systemic velocities were retrieved from the {\it Gaia} DR3 \citep{Gaia2023} database. 10,832 objects were identified as CVs in this catalogue. To do this, we matched our catalogue with the {\it Gaia} DR3 catalogue and found the {\it Gaia} ID for each CV. For these systems, \texttt{phot\_g\_mean\_flux\_over\_error}, $f_{\rm G}/{\delta f_{\rm G}}, > 8$, more than eight “good” astrometric observations, as characterized by {\it Gaia} DR3, \texttt{astrometric\_excess\_noise<2}, and $\varpi>0.1$ mas were considered. Although we have collected $\gamma$ velocities of 455 systems from the literature, only 432 of them have {\it Gaia} data as a result of the above criteria used in matching with the {\it Gaia} DR3 database. We classified CVs in the sample as magnetic (Polars and IPs, hereafter referred to as mCVs) and non-magnetic systems (hereafter referred to as non-mCVs) depending on their classification in the catalogues \citep[e.g.][]{Ritter2003} and their studies in the literature. 

Since the examined systems are located from about 45 pc to 6 kpc as seen in Figure \ref{fig:distances}, the trigonometric parallax data may be affected by some biases. To overcome this issue, \citet{BailerJones2021} recalculated the distances and their errors for all stars with known {\it Gaia} trigonometric parallaxes using Bayesian statistics and a Galaxy model. Thus, we matched our catalogue also with \citet{BailerJones2021} using the {\it Gaia} DR3 IDs to obtain precise distances and distance errors for the 432 CVs in our sample. Figure \ref{fig:distances} shows a comparison of distances calculated from {\it Gaia} DR3 \citep{Gaia2023} database and \citet{BailerJones2021} parallaxes. As can be seen from Figure \ref{fig:distances}, distances found from the two methods are in good agreement up to about 2 kpc, while distances obtained from \citet{BailerJones2021} are smaller than those of {\it Gaia} database for beyond about 2 kpc. As distances of almost all systems in the sample are smaller than 2 kpc and they are estimated from precise trigonometric parallaxes, we concluded that our sample is reliable concerning distances. {\it Gaia} distances of V909 Sgr, CQ Vel and V373 Sct are larger than those estimated by \citet{BailerJones2021}. An inspection shows that $G > 18.5$ mag and $\mid b \mid < 10^{\circ}.5$ for these systems, where $b$ is the Galactic latitude. Thus, disagreement between distances obtained from the two methods for these systems may be due to the use of constant Galactic length scale in \citet{BailerJones2021} \citep[see also][]{Canbay2023}. Indeed, \citet{Bilir2006a} and \citet{YazGokceetal} showed in their star counts for star fields at the same Galactic latitude but different Galactic longitudes that the length scale values of the thin disk are in a wide range and that the length scale is a function of Galactic longitude. 

In this study, to minimize the bias in the trigonometric parallax data of the 432 systems, the distances and distance errors provided by \citet{BailerJones2021} were used. Proper motion components and distances of the CVs in our sample are listed in Table 1. Histograms of relative parallax errors and proper motion errors of the CV sample are shown in Figure \ref{fig:inputerrors}b and \ref{fig:inputerrors}c, respectively, where IPs and Ps are shown separately. As magnetic systems are relatively faint objects, CVs close to the Sun could be detected in general. As magnetic systems tend to be faint, those in the solar neighborhood tend to be predominantly detected. Nearby systems tend to have small parallax errors, though even faint systems beyond a few hundred pc become hard to detect. The median value and standard deviation of relative parallax errors are calculated at 0.028 and 0.070 for all CVs in the sample, respectively. The median value and standard deviation of relative proper motion errors are 0.080 and 0.143 for all CVs, respectively. 
 

\begin{table*}
\label{tab:bigtable}
\setlength{\tabcolsep}{3pt}
\begin{center}
\tiny{
\caption{Equatorial coordinates ($\alpha, \delta$), type (mCV or non-mCV), orbital period ($P_{\rm orb}$), trigonometric parallax ($\varpi$), distance from $\varpi$, 
proper motion components ($\mu_{\alpha}\cos\delta$, $\mu_{\delta}$), and distance from \citet{BailerJones2021} ($d_{\rm BJ}$), and systemic velocity ($\gamma$) of CVs in the sample. The reference for $\gamma$ velocity is given in the last column. The table can be obtained electronically.}
\begin{tabular}{llccccccccccc}
\hline
ID   &  Star Name                               &  $\alpha_{\rm J2000}$   & $\delta_{\rm J2000}$     & Type    & $P_{\rm orb}$ & $\overline{\omega}$    &$d_{\varpi}$               & $\mu_{\alpha}\cos\delta$  & $\mu_{\delta}$          &  $d_{\rm BJ}$          & $\gamma$        & Ref  \\ 
     &                                          & (hh:mm:ss)                & (dd:mm:ss)             &         &     (d)       &     (mas)              &  (pc)                     &(mas yr$^{-1}$)            & (mas yr$^{-1}$)         &    (pc)                 & (km~s$^{-1}$)   &   \\
 \hline
001 & WW Cet                                   		 & 00 11 27.77 & -11 28 43.10     & non-mCV &   0.176     &     4.541$\pm$0.030   &      220$\pm$2     &  12.458$\pm$0.031      &  13.510$\pm$0.022     &  218$\pm$1     &   5.8$\pm$3.7    & (001) \\
002 & 1RXS J001538.2+263656                   & 00 15 38.23 & +26 36 56.81    & non-mCV &   0.102     &     1.804$\pm$0.107    &      554$\pm$33    &  22.390$\pm$0.098      &   -10.733$\pm$0.103   &  558$\pm$35    &   -36$\pm$3  	& (002) \\	
003 & V513 Cas                                  		& 00 18 14.91 & +66 18 13.65     & non-mCV &   0.216     &     1.167$\pm$0.030    &      857$\pm$22    &  6.034$\pm$0.030        &   -1.398$\pm$0.032     &  835$\pm$20    &   -35.9$\pm$0.4  & (003) \\
004 & V592 Cas                                 		& 00 20 52.24 & +55 42 16.20     & non-mCV &   0.115    &     2.145$\pm$0.017    &      466$\pm$4     & -7.133$\pm$0.014        &   -14.501$\pm$0.015   &  460$\pm$3      &   21$\pm$14      & (004) \\
005 & FL Psc                                   		& 00 25 11.11  & +12 17 12.38    & non-mCV &   0.057    &     6.330$\pm$0.103   &      158$\pm$3     &  -70.239$\pm$0.130     &   -37.349$\pm$0.085   & 158$\pm$3       &   -17$\pm$3    & (005) \\
... & ... & ... & ... & ... &  ...  & ...  & ... & ... & ...  & ...& ... & .... \\
... & ... & ... & ... & ... &  ...  & ...  & ... & ... & ...  & ...& ... & .... \\
... & ... & ... & ... & ... &  ...  & ...  & ... & ... & ...  & ...& ... & .... \\
428 & V378 Peg                                		 & 23 40 04.33 & +30 17 47.70   & non-mCV &   0.139   &     1.014$\pm$0.021   &     986$\pm$21      & -5.711$\pm$0.024     &   -5.160$\pm$0.014    & 949$\pm$17     & -3.4$\pm$2    & (139) \\
429 & 1RXS J234015.8+764207                  & 23 40 20.65 & +76 42 10.50   & mCV        &   0.154   &     1.788$\pm$0.052   &     559$\pm$16     & 19.682$\pm$0.054      &   2.412$\pm$0.050    & 552$\pm$15     & -21$\pm$8      & (033) \\
430 & HX Peg                                    		& 23 40 23.69 & +12 37 41.71    & non-mCV &   0.200  &     1.656$\pm$0.031   &     604$\pm$11    	& -14.502$\pm$0.036      &   6.804$\pm$0.023    & 586$\pm$11      &  29$\pm$6      & (140) \\
431 & V630 Cas                                  		& 23 48 51.90 & +51 27 39.10    & non-mCV &   2.564   &     0.381$\pm$0.037    &    2623$\pm$257    & -0.863$\pm$0.032     &   -4.095$\pm$0.031    & 2409$\pm$247   &  -87.3$\pm$2.3   & (141) \\
432 & Gaia 14ade                               		 & 23 50 52.01 & +28 58 59.50   & non-mCV &   0.054   &     0.419$\pm$0.251    &    2389$\pm$1431   & -5.602$\pm$0.236     &  -5.783$\pm$0.142   & 1907$\pm$583    & -1.3$\pm$0.4    & (008) \\
\hline
\end{tabular}  
}
\end{center}
(001) \citet{Ringwald1996}, (002) \citet{Thorstensen2016}, (003) \citet{Szkody2013}, (004) \citet{Huber1998}, (005) \citet{Thorstensen2016}, ..., (008) \citet{Szkody2018}, (133) \citet{Halpern2015}, (139) \citet{Ringwald2012}, (140) \citet{Ringwald1994}, (141) \citet{Orosz2001}
 \end{table*}


\subsection{Galactic Space Velocities}

The space velocities relative to the Sun were calculated using the algorithms and transformation matrices developed by \citet{Johnson1987}. Equatorial coordinates ($\alpha,\delta$), proper-motion components ($\mu_{\alpha} \cos \delta, \mu_{\delta}$), distances \citep[$d_{\rm BJ}$,][]{BailerJones2021}, and systemic velocities ($\gamma$) in Table 1 were used as the main input data. The transformation matrices of \citet{Johnson1987} use the right-handed system notation. Consequently, the space velocity components $U$, $V$, and $W$ represent the velocity vector of a star relative to the Sun. $U$ is directed toward the Galactic Center ($l=0^{o}$, $b=0^{o}$), $V$ towards the Galactic rotation direction ($l=90^{o}$, $b=0^{o}$), and $W$ towards the North Galactic Pole ($b=90^{o}$), where $l$ and $b$ are the Galactic longitude and latitude, respectively.

\begin{figure}
  \centering
  \includegraphics[width=\columnwidth]{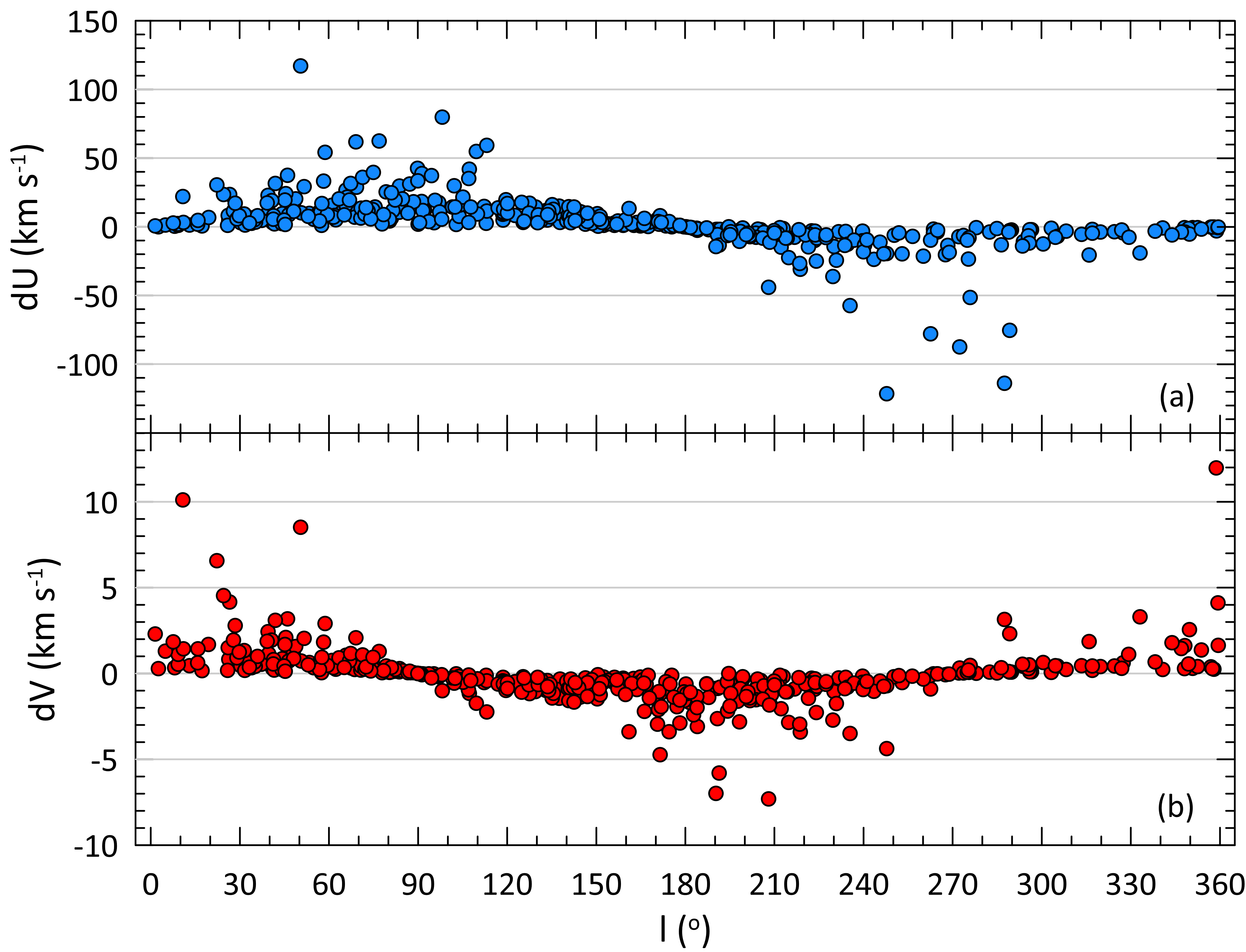}
   \caption{The differential rotation corrections $dU$ (a) and $dV$ (b) applied to the $U$ and $V$ space velocity components of the 432 CVs in the sample as a function of the Galactic longitude .}
 \label{fig:differance_corrections}
\end{figure}

As the distances of CVs in the sample are in a wide range from about 45 pc to 6 kpc, a correction for the differential rotation of the systems around the Galactic centre is required. We applied the procedure of \citet{Mihalas1981} to the distribution of the CVs and estimated the first-order Galactic differential rotation corrections for the $U$ and $V$ space velocity components of the sample. The ranges of these corrections are $-122<dU<118$ and $-8<dV<12$ km s$^{-1}$ for $U$ and $V$ space velocity components, respectively. The $W$ velocity is not affected in this first-order approximation. Variation of the differential correction values applied to the $U$ and $V$ space velocity components of the CVs in the sample with Galactic longitude is shown in Figure \ref{fig:differance_corrections}. As expected, $U$ is affected more than the $V$ component. The high values for the $U$ component show that corrections for differential Galactic rotation cannot be ignored. In addition, the Galactic space velocity components were reduced to the Local Standard of Rest (LSR) by adding the Sun’s space velocity to the velocity components of the CVs. The space velocity of the Sun is adopted as $(U, V, W)_{\odot}=(8.83\pm 0.24, 14.19\pm 0.34, 6.57\pm 0.21)$ km~s$^{-1}$ \citep{Coskunoglu2011}. Galactic space velocity components corrected for the LSR are listed in Table 2.

\begin{figure}
  \centering
  \includegraphics[width=\columnwidth]{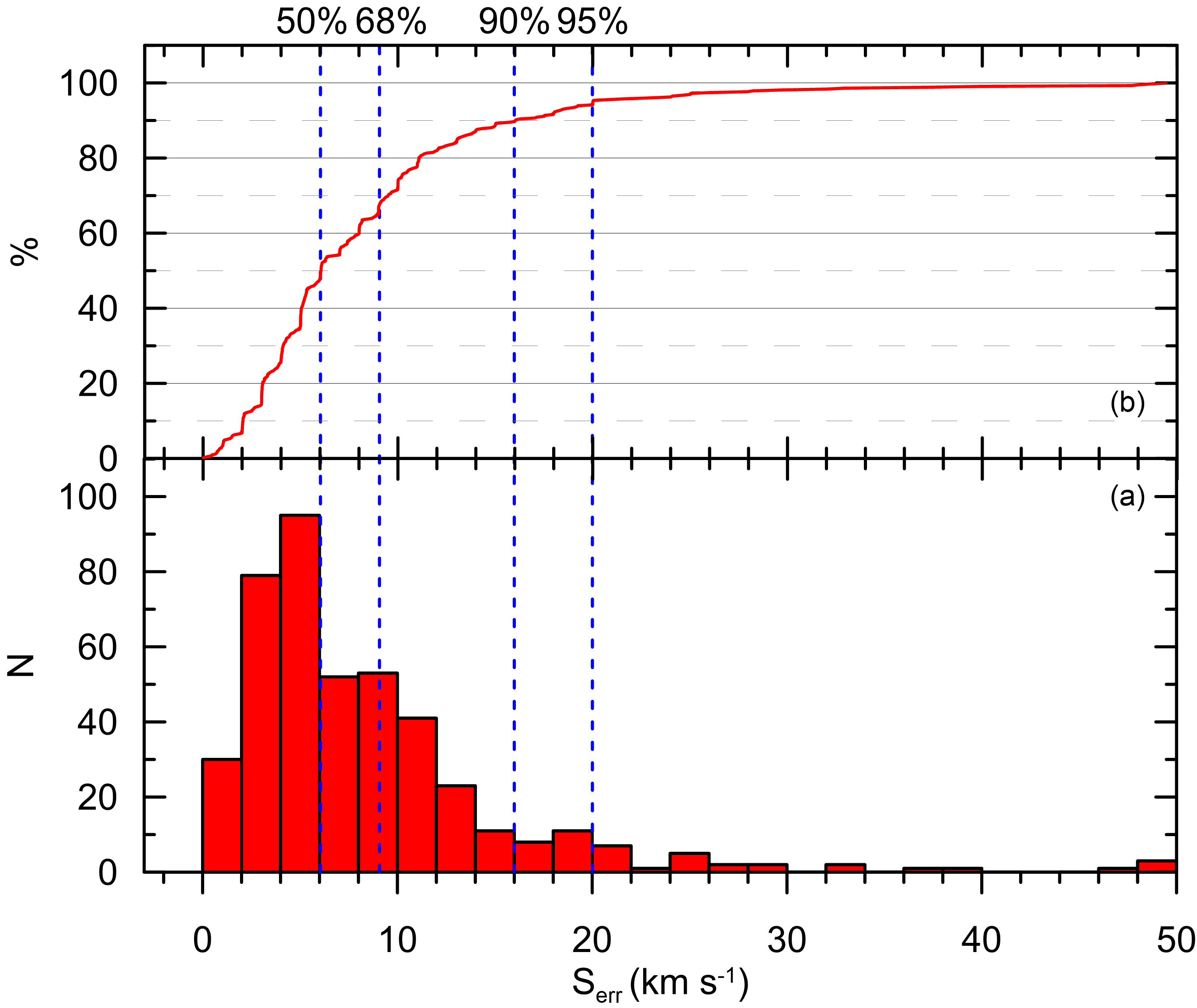}
   \caption{(a) Cumulative distribution and (b) histogram of the total space velocity errors of 432 CVs in the sample.}
   \label{fig:space_err}
\end{figure}

Uncertainties of the space velocity components were determined through the propagation of uncertainties from the input data, employing the algorithm provided by \citet{Johnson1987}. Uncertainties of the total space velocities $S_{\rm err}=\sqrt{U_{\rm err}^2+V_{\rm err}^2+W_{\rm err}^2}$ were also computed. A histogram depicting the total space velocity errors of 432 CVs was constructed and shown in Figure \ref{fig:space_err}, where their cumulative distribution was also displayed. We determined from this histogram that the total space velocity errors corresponding to 50\%, 68\%, 90\% and 95\% of the sample are $S_{\rm err} = 6.05$ km~s$^{-1}$, $S_{\rm err} = 9.08$ km~s$^{-1}$, $S_{\rm err} = 16.02$ km~s$^{-1}$ and $S_{\rm err} = 20.01$ km~s$^{-1}$, respectively. 

In order to increase the reliability of space velocities to indicate kinematical ages, the dispersion is expected to be larger than the propagated errors. Mean values and dispersions of the space velocity components and their dispersions, total space velocity dispersions, and kinematical ages for the CV groups classified according to the total space velocity error are listed in Table 3. The space velocity dispersion of the sample increases as stars with large space velocity errors are taken into account, which leads to unreliable calculation of group ages. The $U-V$ and $W-V$ diagrams of the four groups separated according to their total space velocity errors are shown in Figure \ref{fig:UVW_err}a and b, while Figure \ref{fig:UVW_err}c and d displays the $U-V$ and $W-V$ diagrams of non-magnetic systems, polars and IPs. Figure \ref{fig:UVW_err}a and b reveals that systems with large space velocity errors have larger space velocity components, while magnetic systems were mostly found in lower space velocity regions of the diagrams compared to non-magnetic systems. We removed systems with $S_{\rm err} > 16.02$ km~s$^{-1}$ to increase the reliability of the analysis, reducing the number of CVs in the final sample to 385.

\begin{table*}
\centering
\setlength{\tabcolsep}{2.5pt}
\scriptsize{
\caption{Sun-centred rectangular Galactic coordinates ($X$, $Y$, and $Z$), space velocity components corrected for the LSR ($U_{\rm LSR}$, $V_{\rm LSR}$, and $W_{\rm LSR}$) and $TD/D$ values for CVs in the sample. The table can be obtained electronically.}
\begin{tabular}{llcccccccccc}
\hline
ID   &  Star Name &  $l$   &   $b$   &  $d_{\rm BJ}$   &$X$   &   $Y$   &  $Z$   &  $U_{\rm LSR}$    & $V_{\rm LSR}$   &  $W_{\rm LSR}$  & $TD/D$    \\ 
     &            &  ($^{\rm o}$)  &   ($^{\rm o}$)  &  (pc)  &(pc)  &   (pc)  &  (pc)  &  (km~s$^{-1}$)    &  (km~s$^{-1}$)  &  (km~s$^{-1}$)  &           \\
 \hline
001 & WW Cet                                    &90.01  & -71.74             &  218$\pm$1      	&   0      &    68     &  -207     	&  -11.12$\pm$0.09          &   21.50$\pm$1.16         &  2.87$\pm$3.51             &   0.009     \\
002 & 1RXS J001538.2+263656        & 113.09  & -35.57           &  558$\pm$35  	 	&   -178     &     418     & -325    	&  -31.61$\pm$2.60          &   -53.68$\pm$3.32         &  -3.45$\pm$2.54           &   0.070  \\
003 & V513 Cas                                  & 119.60 & 3.65               &  835$\pm$20 		&  -411     &      724     &  53    	 &  -13.41$\pm$0.52         &   -26.84$\pm$0.43          &  -4.26$\pm$0.24            &   0.009     \\
004 & V592 Cas                                 & 118.60  & -6.91              &  460$\pm$3 		&  -219      &     401     & -55     	& 6.51$\pm$6.65         &   39.09$\pm$12.20         &  -25.28$\pm$1.70           &   0.069       \\
005 & FL Psc                                   & 112.91& -50.07                & 158$\pm$3 		&  -39      &    93     &  -121     	&  68.23$\pm$1.31          &   12.56$\pm$1.78         & 7.23$\pm$2.31            &   0.027    \\

...  & ... & ... & ... &... & ... & ... & ... &  ...  & ...  & ... & ...   \\
...  & ... & ... & ... &... & ... & ... & ... &  ...  & ...  & ... & ...   \\
...  & ... & ... & ... &... & ... & ... & ... &  ...  & ...  & ... & ...   \\
428 & V378 Peg                                    &105.13 & -30.10                & 949$\pm$17 		&  -214      &    792     &     -476     	&  20.46$\pm$0.75          &   13.39$\pm$1.67         &  -3.78$\pm$1.03             &   0.007     \\
429 &1RXS J234015.8+764207           & 118.77  & 14.41       	& 552$\pm$15 		&   -257      &     469     &      137    	&  -40.00$\pm$3.92         &   -25.83$\pm$6.82         &  -6.78$\pm$2.01           &   0.014  \\
430 & HX Peg                                       & 97.23 & -46.67                  & 586$\pm$11 		&   -51     &      399     &      -426     	&  21.73$\pm$0.71          &   61.80$\pm$4.12          &  8.33$\pm$4.38            &   0.293    \\
431 & V630 Cas                                  & 113.10  & -10.21                 & 2409$\pm$247 		&  -930      &      2181     &     -427     	&  5.49$\pm$2.25         &   -61.39$\pm$2.09         &  -20.23$\pm$3.91           &   0.207       \\
432 & Gaia 14ade                                & 107.29  & -32.04            	& 1907$\pm$583 		&  -481      &     1545     &     -1012     	&  33.12$\pm$15.48          &   14.60$\pm$1.10        & -23.60$\pm$7.28            &   0.022    \\
\hline
\end{tabular}  
}
 \end{table*}


\begin{table*}
\setlength{\tabcolsep}{2.2pt}
\renewcommand{\arraystretch}{1.4}
\begin{center}
\scriptsize{
\caption{Mean space velocity components ($\langle U_{\rm LSR}\rangle$, $\langle V_{\rm LSR}\rangle$ and $\langle W_{\rm LSR}\rangle$) with their errors, space velocity dispersions ($\sigma_{U}$, $\sigma_{V}$ and $\sigma_{W}$), total space velocity dispersions ($\sigma_{\nu}$), kinematical ages ($\tau$) and $\gamma$ velocity dispersions ($\sigma_{\gamma}$) for CV groups. $N$ is the number of systems.}
\begin{tabular}{lcccccccccc}
\hline
Group & $N$ & $\langle U_{\rm LSR}\rangle$ &  $\langle V_{\rm LSR}\rangle$ & $\langle W_{\rm LSR}\rangle$ & $\sigma_{U{_{\rm LSR}}}$ & $\sigma_{V{_{\rm LSR}}}$ & $\sigma_{W{_{\rm LSR}}}$ & $\sigma_{\nu}$ & $\tau$ & $\sigma_{\gamma}$  \\ 
     &   & (km~s$^{-1}$)  &  (km~s$^{-1}$)  & (km~s$^{-1}$)  &  (km~s$^{-1}$) &  (km~s$^{-1}$) & (km~s$^{-1}$) &  (km~s$^{-1}$)  &    (Gyr)   &  (km~s$^{-1}$)   \\
 \hline
 \multicolumn{11}{c}{\bf Preliminary sample of 432 CVs}\\
  \hline
$S_{\rm err}\leq$ 6.05 (km~s$^{-1}$)  & 214 & 0.76$\pm$1.98 & -5.46$\pm$2.09 & -3.28$\pm$1.60 & 35.47$\pm$1.24 & 29.35$\pm$1.35 & 21.02$\pm$1.12 & 50.61$\pm$2.15 & 4.72$\pm$0.39 & 29.22$\pm$1.24 \\
$S_{\rm err}\leq$ 9.08  (km~s$^{-1}$) & 291 & 0.32$\pm$2.59 & -5.25$\pm$2.71 & -2.16$\pm$2.05 & 36.15$\pm$1.80 & 29.89$\pm$1.84 & 22.37$\pm$1.48 & 51.97$\pm$2.97 & 4.97$\pm$0.54 & 30.00$\pm$1.72 \\
$S_{\rm err}\leq$ 16.02 (km~s$^{-1}$) & 385 & 1.08$\pm$3.65 & -6.52$\pm$3.55 & -2.75$\pm$2.78 & 38.66$\pm$2.80 & 29.90$\pm$2.67 & 23.64$\pm$2.31 & 54.29$\pm$4.51 & 5.40$\pm$0.83 & 31.34$\pm$2.61 \\
$S_{\rm err}\leq$ 20.01 (km~s$^{-1}$) & 407 & 1.61$\pm$3.98 & -6.97$\pm$3.94 & -2.88$\pm$3.06 & 39.52$\pm$3.24 & 31.36$\pm$3.27 & 23.63$\pm$2.75 & 55.71$\pm$5.36 & 5.66$\pm$0.98 & 32.16$\pm$3.10 \\
All Systems                           & 432 & 2.30$\pm$5.01 & -10.01$\pm$4.98& -3.70$\pm$3.68 & 44.03$\pm$5.85 & 40.79$\pm$7.62 & 26.22$\pm$5.02 & 65.50$\pm$10.84 & 7.43$\pm$1.92 & 37.82$\pm$6.25 \\
\hline
\hline
 \multicolumn{11}{c}{\bf Final sample of 385 CVs with $S_{\rm err} \leq 16.02$ km~s$^{-1}$}\\
\hline
 $TD/D\leq0.1$    & 307 & -0.74$\pm$0.53 & -4.46$\pm$3.44 & -0.33$\pm$2.54 & 30.84$\pm$2.74 & 20.47$\pm$2.63 & 15.72$\pm$2.18 & 40.21$\pm$4.38 & 2.89$\pm$0.72 & 23.22$\pm$2.52 \\
$0.1<TD/D\leq1$  & ~~47 & 10.47$\pm$4.14 & -10.19$\pm$4.04 & -6.18$\pm$3.31 & 50.78$\pm$3.34 & 40.71$\pm$2.65 & 30.03$\pm$2.27 & 71.68$\pm$4.83 & 8.51$\pm$0.83 & 41.38$\pm$2.79 \\
$1<TD/D\leq10$   & ~~13 & -9.84$\pm$3.46 & -18.54$\pm$4.55 & -23.86$\pm$3.52 & 56.38$\pm$2.44 & 50.22$\pm$3.85 & 34.33$\pm$2.19 & 82.94$\pm$5.06 & 10.37$\pm$0.80 & 47.89$\pm$2.92 \\
$TD/D>10$        & ~~18 & 15.39$\pm$4.52 & -21.49$\pm$3.63 & -21.15$\pm$5.19 & 78.33$\pm$2.25 & 74.93$\pm$2.37 & 61.67$\pm$2.89 & 124.71$\pm$4.36 & 16.03$\pm$0.50 & 72.00$\pm$2.52 \\
All CVs        & 385 & 1.08$\pm$3.65 & -6.52$\pm$3.55 & -2.75$\pm$2.78 & 38.66$\pm$2.80 & 29.90$\pm$2.67 & 23.64$\pm$2.31 & 54.29$\pm$4.51 & 5.40$\pm$0.83 & 31.34$\pm$2.61 \\
\hline
$TD/D\leq1$    & 354 & 0.75$\pm$3.61 & -5.22$\pm$3.52 & -1.10$\pm$2.64 & 34.37$\pm$2.83 & 24.23$\pm$2.64 & 18.39$\pm$2.21 & 45.90$\pm$4.46 & 3.87$\pm$0.79 & 26.50$\pm$2.58 \\
\hline
 \hline
 \multicolumn{11}{c}{\bf Final sample of 354 CVs with $S_{\rm err} \leq 16.02$ km~s$^{-1}$ and $TD/D \leq 1$}\\
\hline
$P_{\rm orb}<2.15$ (h) & 139 & 0.39$\pm$3.62 & -7.74$\pm$3.50 & -1.06$\pm$2.98 & 35.09$\pm$2.61 & 25.22$\pm$2.54 & 19.58$\pm$2.07 & 47.44$\pm$4.19 & 4.15$\pm$0.75 & 27.39$\pm$2.42 \\
$P_{\rm orb}>3.18$ (h) & 188 & 1.16$\pm$3.54 & -3.21$\pm$3.39 & -2.13$\pm$2.31 & 33.45$\pm$2.93 & 22.83$\pm$2.51 & 16.49$\pm$2.25 & 43.73$\pm$4.47 & 3.49$\pm$0.78 & 25.25$\pm$2.58 \\
\hline
mCVs              & ~~49  & 0.71$\pm$5.05 & -6.38$\pm$4.20 & 4.31$\pm$3.18 & 33.64$\pm$3.48 & 21.84$\pm$2.96 & 14.60$\pm$2.82 & 42.68$\pm$5.37 & 3.31$\pm$0.92 & 24.64$\pm$3.10 \\
non-mCVs          & 305 & 0.76$\pm$3.38 & -5.03$\pm$3.41 & -1.98$\pm$2.56 & 34.49$\pm$2.64 & 24.59$\pm$2.57 & 18.78$\pm$2.08 & 46.33$\pm$4.23 & 3.95$\pm$0.75 & 26.75$\pm$2.44 \\
\hline
$P_{\rm orb}<2.15$ (h) (non-mCVs) & 123 & 1.06$\pm$3.42 & -7.67$\pm$3.47 & -1.98$\pm$2.95 & 34.90$\pm$2.41 & 25.98$\pm$2.43 & 19.47$\pm$1.96 & 47.67$\pm$3.94 & 4.19$\pm$0.71 & 27.52$\pm$2.28 \\
$P_{\rm orb}>3.18$ (h) (non-mCVs) & 161 & 0.75$\pm$3.30 & -2.95$\pm$3.20 & -2.73$\pm$2.19 & 33.92$\pm$2.79 & 22.97$\pm$2.38 & 17.20$\pm$2.09 & 44.43$\pm$4.22 & 3.61$\pm$0.74 & 25.65$\pm$2.44 \\
\hline
$0.030<P_{\rm orb} (\rm d)\leq0.072$ & 77 & -3.71$\pm$3.59 & -10.57$\pm$3.32 & -0.30$\pm$2.54 & 36.44$\pm$2.61 & 25.05$\pm$2.48 & 20.18$\pm$1.83 & 48.61$\pm$4.04 & 4.36$\pm$0.73 & 28.06$\pm$2.34 \\
$0.072<P_{\rm orb} (\rm d)\leq0.110$ & 75 & 4.84$\pm$3.63 & -3.33$\pm$3.97 & -0.45$\pm$3.53 & 33.25$\pm$2.58 & 25.60$\pm$2.89 & 19.35$\pm$2.16 & 46.21$\pm$4.44 & 3.93$\pm$0.79 & 26.68$\pm$2.56 \\
$0.110<P_{\rm orb} (\rm d)\leq0.170$ & 86 & -0.45$\pm$3.77 & -6.43$\pm$3.50 & -1.62$\pm$2.55 & 33.03$\pm$3.18 & 24.03$\pm$2.80 & 20.12$\pm$2.54 & 45.53$\pm$4.94 & 3.81$\pm$0.87 & 26.29$\pm$2.85 \\
$0.170<P_{\rm orb} (\rm d)\leq0.280$ & 70 & 3.20$\pm$3.48 & -3.33$\pm$3.58 & -2.98$\pm$2.37 & 35.02$\pm$2.85 & 23.18$\pm$2.46 & 14.04$\pm$2.18 & 44.28$\pm$4.35 & 3.59$\pm$0.76 & 25.57$\pm$2.51 \\
$0.280<P_{\rm orb} (\rm d)\leq0.370$ & 24 & 0.53$\pm$3.37 & 3.57$\pm$2.76 & 5.28$\pm$1.81 & 35.78$\pm$2.45 & 18.50$\pm$1.60 & 11.62$\pm$1.59 & 41.92$\pm$3.33 & 3.18$\pm$0.57 & 24.20$\pm$1.93 \\
$0.370<P_{\rm orb} (\rm d)\leq0.550$ & 14 & -0.37$\pm$3.56 & 0.87$\pm$3.70 & -7.76$\pm$1.86 & 31.53$\pm$3.57 & 16.75$\pm$2.94 & 14.18$\pm$2.09 & 38.42$\pm$5.08 & 2.60$\pm$0.81 & 22.18$\pm$2.93 \\
$0.550<P_{\rm orb} (\rm d)$          & 08 & -0.43$\pm$3.85 & -11.87$\pm$2.94 & -0.50$\pm$2.56 & --- & --- & ---& --- & --- & --- \\
\hline
\hline
 \multicolumn{11}{c}{\bf Final sample Non-magnetic CVs with $S_{\rm err} \leq 16.02$ km~s$^{-1}$}\\
\hline
 $TD/D\leq0.1$  & 263  & -0.60$\pm$3.32 & -4.52$\pm$3.33 & -1.27$\pm$2.2.46& 30.87$\pm$2.56 & 20.82$\pm$2.53 & 15.94$\pm$2.04 & 40.50$\pm$4.14 & 2.94$\pm$0.69 & 23.38$\pm$2.39 \\
$0.1<TD/D\leq1$ & ~~42 &  9.25$\pm$3.77 & -8.23$\pm$3.91 & -6.37$\pm$3.18 & 50.87$\pm$3.11 & 40.79$\pm$2.73 & 30.77$\pm$2.22 & 72.10$\pm$4.70 & 8.58$\pm$0.80 & 41.63$\pm$2.71 \\
$1<TD/D\leq10$  & ~~8  & -8.69$\pm$4.02 & -48.60$\pm$4.87 & -11.62$\pm$3.67 & 65.65$\pm$2.52 & 23.11$\pm$3.84 & 37.64$\pm$1.77 & 79.12$\pm$5.00 & 9.76$\pm$0.81 & 45.68$\pm$2.89 \\
$TD/D>10$       & ~~15 & 17.63$\pm$4.83 & -14.14$\pm$3.49 & -22.78$\pm$5.04 & 82.03$\pm$2.33 & 78.64$\pm$2.19 & 61.41$\pm$2.57 & 129.17$\pm$4.10 & 16.54$\pm$0.46 & 74.58$\pm$2.36 \\
All non-mCVs   & 328  & 1.30$\pm$3.46 & -6.51$\pm$3.45 & -3.16$\pm$2.70 & 39.17$\pm$2.65 & 30.10$\pm$2.61 & 23.57$\pm$2.17 & 54.73$\pm$4.31 & 5.48$\pm$0.79 & 31.60$\pm$2.49 \\
\hline
\hline
 \multicolumn{11}{c}{\bf Final sample Non-Magnetic CVs with $S_{\rm err} \leq 16.02$ km~s$^{-1}$ and $TD/D \leq 1$}\\
 \hline
$0.030<P_{\rm orb} (\rm d)\leq0.072$ & 66 & -4.05$\pm$3.24 & -10.92$\pm$3.06 & -1.51$\pm$2.54 & 35.53$\pm$2.35 & 25.64$\pm$2.22 & 20.10$\pm$1.69 & 48.21$\pm$3.65 & 4.29$\pm$0.66 & 
27.83$\pm$2.11 \\
$0.072<P_{\rm orb} (\rm d)\leq0.110$ & 69 & 5.27$\pm$3.60 & -3.32$\pm$4.13 & -0.93$\pm$3.42 & 33.46$\pm$2.48 & 26.39$\pm$2.89 & 19.47$\pm$2.11 & 46.85$\pm$4.35 & 4.04$\pm$0.78 & 27.05$\pm$2.51 \\
$0.110<P_{\rm orb} (\rm d)\leq0.170$ & 72 & 0.20$\pm$3.49 & -5.54$\pm$3.20 & -3.16$\pm$2.25 & 33.40$\pm$3.06 & 23.71$\pm$2.71 & 20.88$\pm$2.15 & 45.97$\pm$4.62 & 3.88$\pm$0.82 & 26.54$\pm$2.67 \\
$0.170<P_{\rm orb} (\rm d)\leq0.280$ & 57 & 2.90$\pm$3.17 & -3.02$\pm$3.50 & -3.99$\pm$2.29 & 35.37$\pm$2.51 & 23.72$\pm$2.52 & 14.49$\pm$2.23 & 44.98$\pm$4.20 & 3.71$\pm$0.74 & 25.97$\pm$2.42 \\
$0.280<P_{\rm orb} (\rm d)\leq0.370$ & 22 & 0.52$\pm$3.41 & 3.31$\pm$2.78 & 5.26$\pm$1.86 & 37.32$\pm$2.45 & 18.46$\pm$1.66 & 12.14$\pm$1.63 & 43.37$\pm$3.38 & 3.43$\pm$0.59 & 25.04$\pm$1.95 \\
$0.370<P_{\rm orb} (\rm d)\leq0.550$ & 12 & -0.25$\pm$2.82 & 2.13$\pm$3.39 & -8.31$\pm$1.92 & 33.89$\pm$3.11 & 15.23$\pm$2.25 & 14.35$\pm$2.22 & 39.83$\pm$4.43 & 2.83$\pm$0.73 & 23.00$\pm$2.55 \\
$0.550<P_{\rm orb} (\rm d)$          & 07 & -7.66$\pm$4.00 & -16.10$\pm$3.25 & 0.10$\pm$2.83 & --- & --- & ---& --- & --- & --- \\
\hline
\end{tabular}  
}
\end{center}
\end{table*}

\subsection{Velocity Dispersions and Kinematical Ages}

Estimating the kinematical age of subsamples extracted from a large sample is important for determining the kinematical properties of a group of objects. The basis of kinematic age determination lies in calculating reliable space velocity dispersions and systemic velocity dispersions of the star groups under investigation. By determining the group space velocity dispersions, age can be determined using the age-space velocity dispersion relationship. In this study, the age-space velocity dispersion relationship described by \citet{Cox2000} is utilized:

\begin{equation}
\sigma_{\nu}^3 (\tau) = \sigma_{v,\tau=0}^3 + \frac{3}{2} \sigma_V \delta_2 T_\delta \left[ \exp\left(\frac{\tau}{T_\delta}\right) - 1 \right] 
\end{equation}
Here, $\sigma_{\nu,\tau=0}$ represents the total space velocity dispersion at zero age and is generally assumed to be 10 km~s$^{-1}$ \citep{Cox2000}. $\sigma_{\rm V}$ describes the rotation curve, typically taken as $\approx$2.95. $T_{\delta}$ is a timescale of 5 Gyr, and $\delta_{2}$ is the diffusion coefficient of $3.7 \times 10^{-6}$  km~s$^{3}$~yr$^{-1}$. $\sigma_{\rm {\nu}}$ and $\tau$ are the total space velocity dispersion and the kinematic age of the considered CV group, respectively. The relation between the total dispersion of space velocity vectors $\sigma_{\rm {\nu}}$ and the dispersion of space velocity components is described as:

\begin{equation}
\sigma_{\nu}^2 = \sigma_{\rm U}^2 + \sigma_{\rm V}^2 + \sigma_{\rm W}^2 
\end{equation}
After calculating $\sigma_{\nu}^2$ from the dispersions of velocity components using Equation (2), the kinematic age of a certain CV group can be determined using Equation (1). Additionally, under the assumption that CVs are isotropically distributed, the $\gamma$ velocity dispersion $\sigma_{\gamma}$ can also be calculated from the definition $\sigma_\gamma^2 = \frac{1}{3} \sigma_{\nu}^2$ \citep{Wielen1992}, allowing for comparison with theoretical predictions. Total space velocity dispersions, kinematic ages and $\gamma$ velocity dispersions of CV groups are listed in Table 3.


\subsection{Spatial Distributions}

Distributions of the systems in the final sample of 385 CVs according to equatorial and Galactic coordinates are shown in Figures \ref{fig:coordinates}a and \ref{fig:coordinates}b. There are 278 and 107 CVs with positive and negative declination values, respectively, while the numbers of systems with $b\geq 0^{\circ}$ and $b<0^{\circ}$ are 221 and 164, respectively. 

We computed Sun-centred rectangular Galactic coordinates ($X$ towards Galactic Center, $Y$ Galactic rotation, $Z$ North Galactic Pole) to investigate the spatial distribution of the CV sample. Figure \ref{fig:XYZ} displays the projected positions on the Galactic plane ($X, Y$) and the Galactic plane perpendicular to it ($X, Z$).  Polars and IPs are shown separately in Figure \ref{fig:XYZ}. Median values and standard deviations of the $X$, $Y$ and $Z$ coordinates are -82$\pm$503, 104$\pm$509 and 50$\pm$331 pc for all CVs in the final sample, respectively, -82$\pm$504, 111$\pm$461 and 73$\pm$337 pc for non-magnetic CVs, and -97$\pm$500, 58$\pm$727 and 2$\pm$287 pc for magnetic CVs.  These values show that magnetic systems have a smaller dispersion in their spatial distribution. Figure \ref{fig:XYZ} reveals that most of the systems (85\%) are located closer than about 1 kpc. The median distance of the final sample is 570 pc. These median values reveal that the sample is well within the Galactic disk in the solar neighborhood. Figure \ref{fig:XYZ} also shows that magnetic systems are located near the Sun, in general. This is expected since only close magnetic CVs can be detected, as they are relatively faint systems.

\begin{figure*}
  \centering
  \includegraphics[width=2 \columnwidth]{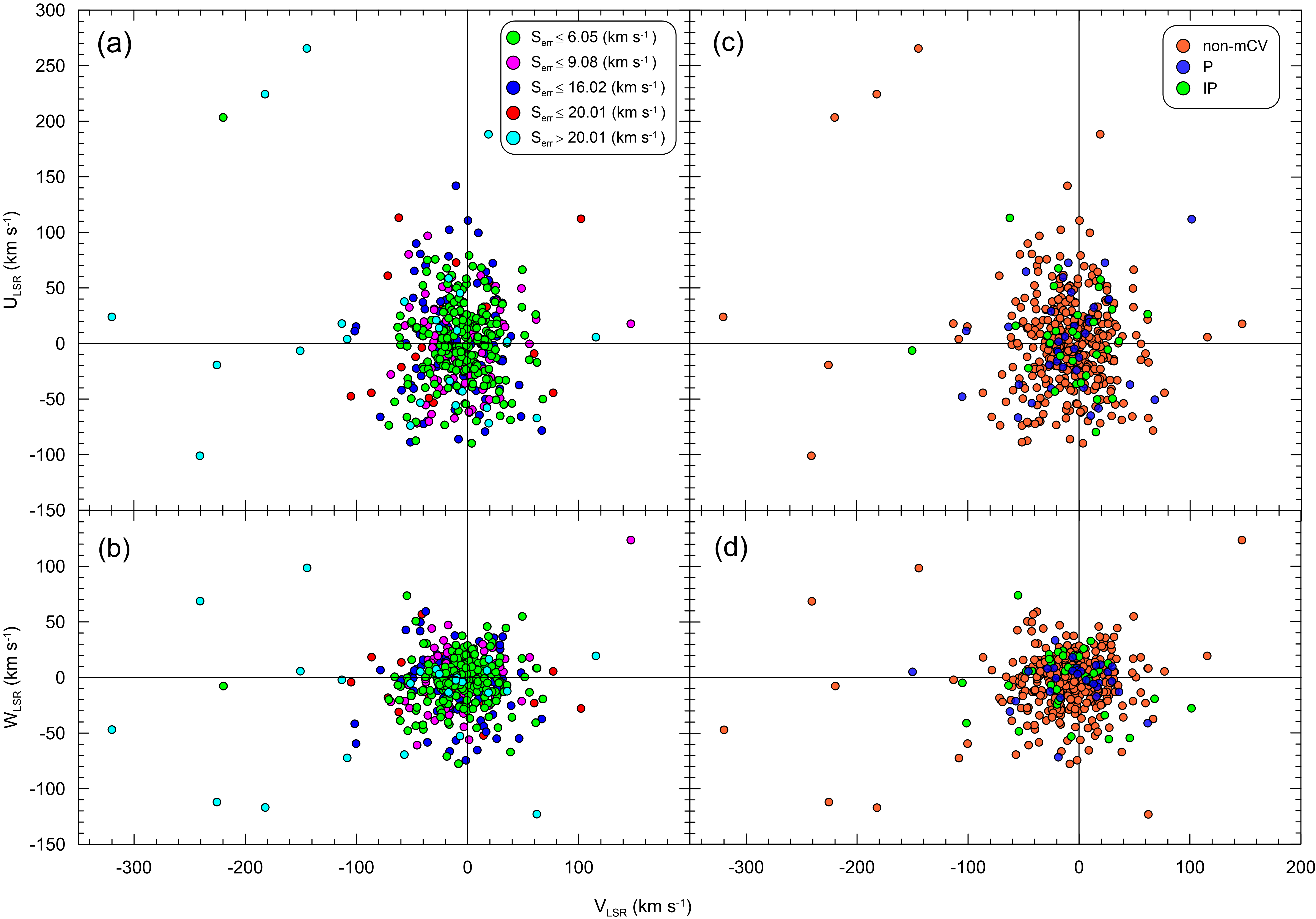}
   \caption{$U-V$ and $W-V$ diagrams of CV groups according to their total space velocity errors $S_{\rm err}$ (a and b), and $U-V$ and $W-V$ diagrams of magnetic (polars (P) and intermediate polars (IP) and non-magnetic systems in the sample of 432 CVs (c and d).}
   \label{fig:UVW_err}
\end{figure*}

\begin{figure}
  \centering
  \includegraphics[width=\columnwidth]{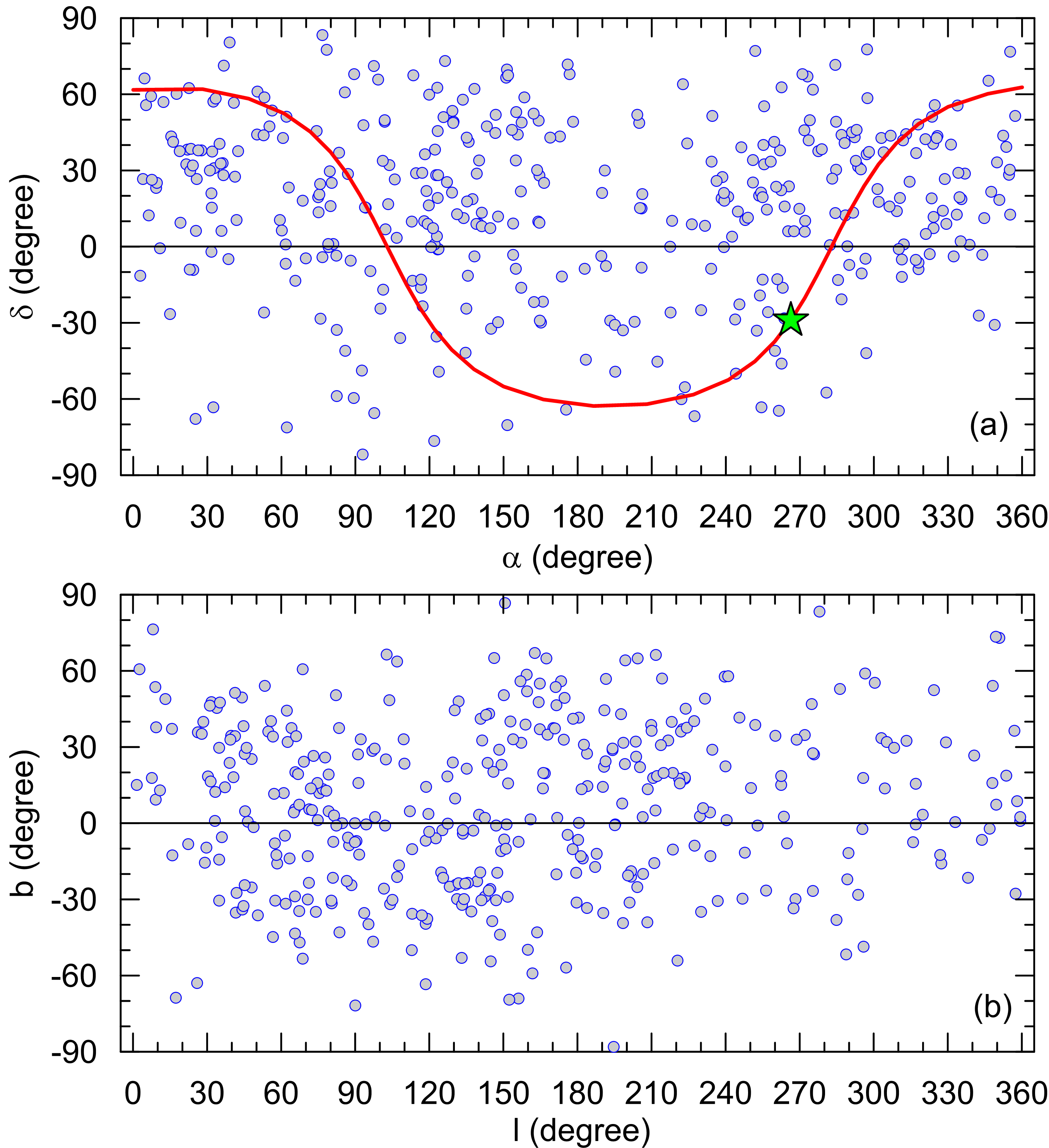}
   \caption{Distribution of 385 CVs in the sample according to their equatorial (a) and Galactic (b) coordinates. In the upper panel, the red curve indicates the Galactic plane and the green star symbol indicates the Galactic center.}
   \label{fig:coordinates}
\end{figure}

\begin{figure}
  \centering
  \includegraphics[width=\columnwidth]{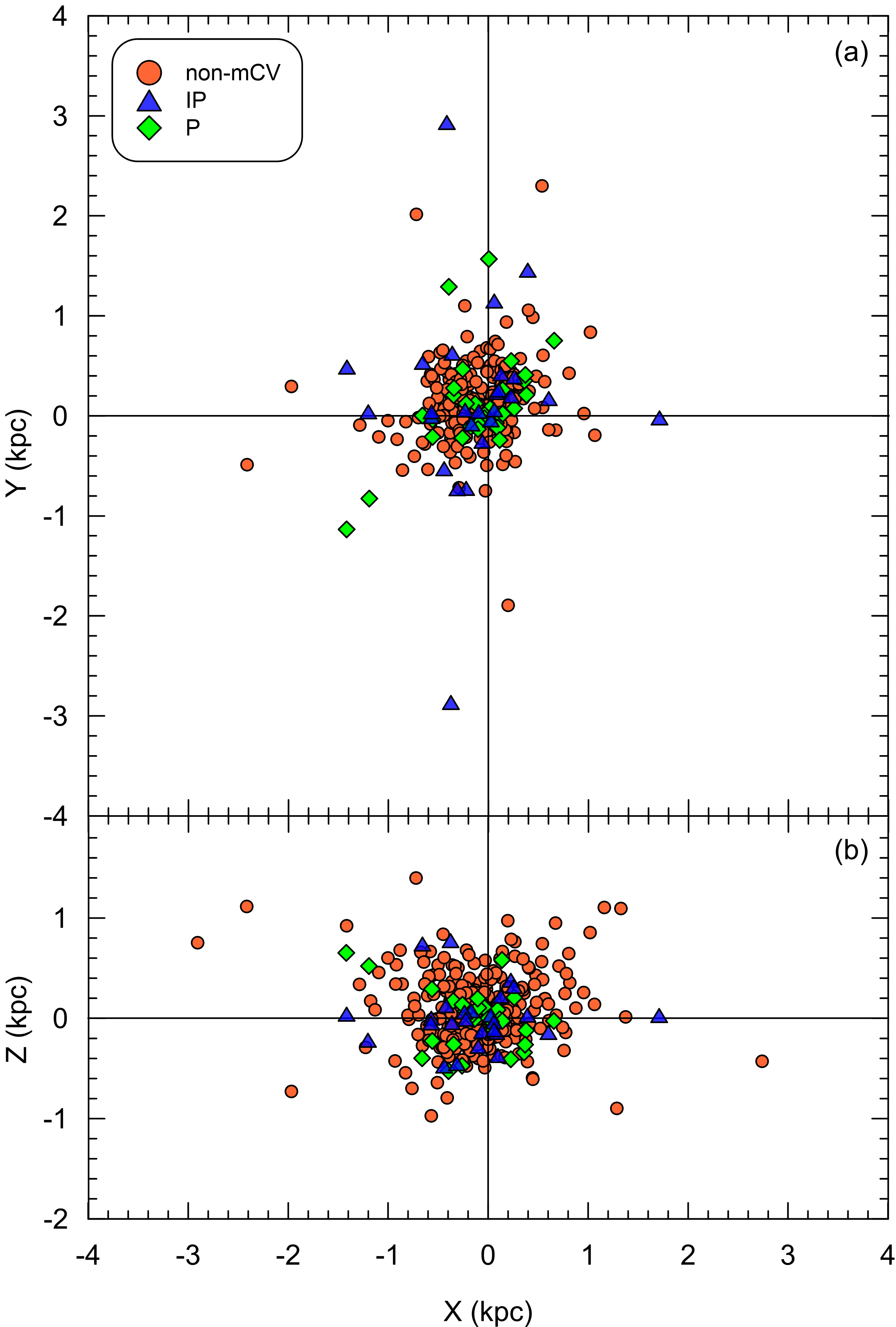}
   \caption{The spatial distribution of 385 CVs in the sample with respect to the Sun. $X$, $Y$ and $Z$ are Sun-centred rectangular Galactic coordinates. P and IP denote polars and intermediate polars, respectively.}
   \label{fig:XYZ}
\end{figure}


\subsection{Population Types}

As the velocity dispersions of object groups from various components of the Galaxy can bias the results, it is crucial to determine which Galactic component an object belongs to. A method for distinguishing between Galactic population classes based on kinematic data was proposed by \citet{Bensby2003, Bensby2005}. Their kinematic criteria were used to determine the Galactic population types of the CVs in our sample. This method assumes that the space velocity components relative to the Local Standard of Rest (LSR), $(U, V, W)_{\rm LSR}$, exhibit a Gaussian-like distribution for the thin disk, thick disk, and halo populations in the Galaxy:

\begin{equation}
P_i(U, V, W) = k \exp \left( -\frac{U_{\text{LSR}}^2}{2\sigma_{i,U}^2} - \frac{(V_{\text{LSR}} - v_{i,a})^2}{2\sigma_{i,V}^2} - \frac{W_{\text{LSR}}^2}{2\sigma_{i,W}^2} \right)  
\end{equation}
where 
\begin{equation}
k =\frac{1}{(2\pi)^{3/2} \sigma_{i,U} \sigma_{i,V} \sigma_{i,W}} 
\end{equation}
$\sigma_U$, $\sigma_V$, and $\sigma_W$ are the velocity dispersions, with values for the thin disk ($i = D$) being 35, 20, and 16 km~s$^{-1}$, respectively; for the thick disk ($i = TD$), 67, 38, and 35 km~s$^{-1}$; and for the halo ($i = H$), 160, 90, and 90 km~s$^{-1}$ \citep{Bensby2003}. The asymmetric drift $\nu_a$ values for the thin disk, thick disk, and halo populations are -15, -46, and -220 km~s$^{-1}$, respectively \citep{Bensby2003}.

To determine whether the systems belong to a particular Galactic population, the probabilities given by Equation 3 are calculated and multiplied by the population fractions ($X$) near the Sun. $X$ represents the population ratios of stars near the Sun, with values for the thin disk, thick disk, and halo being $X_{\rm D}=0.9385$, $X_{\rm TD}=0.06$, and $X_{H}=0.0015$, respectively \citep[e.g.,][]{Buser1999, Bilir2006b, Bilir2006c, Bilir2008, Cabrera-Lavers2007}. The relative probabilities of the thick disk compared to the thin disk and halo are determined by the following equations:

\begin{equation} 
\frac{T D}{D} = \frac{X_{TD}}{X_D} \times \frac{P_{TD}}{P_D}~~~~~~
\frac{TD}{H} = \frac{X_{TD}}{X_H} \times \frac{P_{TD}}{P_H}
\end{equation}
\citet{Bensby2003} classified stars with $TD/D\leq0.1$ as high-probability thin disk stars. According to this classification, the stars selected have a likelihood of belonging to the thin disk that is 10 times higher than the likelihood of belonging to the thick disk. Conversely, stars with $TD/D>10$ are considered high-probability thick disk stars. Stars within the ranges $0.1<TD/D\leq1$ and $1<TD/D\leq10$ are classified as low-probability thin disk and low-probability thick disk stars, respectively.

For 385 CVs in the final sample, the results based on the kinematic population separation by \citet{Bensby2003, Bensby2005} indicate that 307 systems fall within the $TD/D\leq 0.1$ range, 47 systems within the $0.1<TD/D\leq 1$ range, 13 systems within the $1<TD/D \leq 10$ range, and 18 systems within the $TD/D>10$ range (Table 3). Population types  ($TD/D$ values) of CVs in the sample are listed in Table 2. This classification shows that the percentage of high and low probability thin disk CVs in the final sample is 92\%, while 8\% of the final sample are the high- and low-probability thick disk CVs. 

\begin{figure}
  \centering
  \includegraphics[width=\columnwidth]{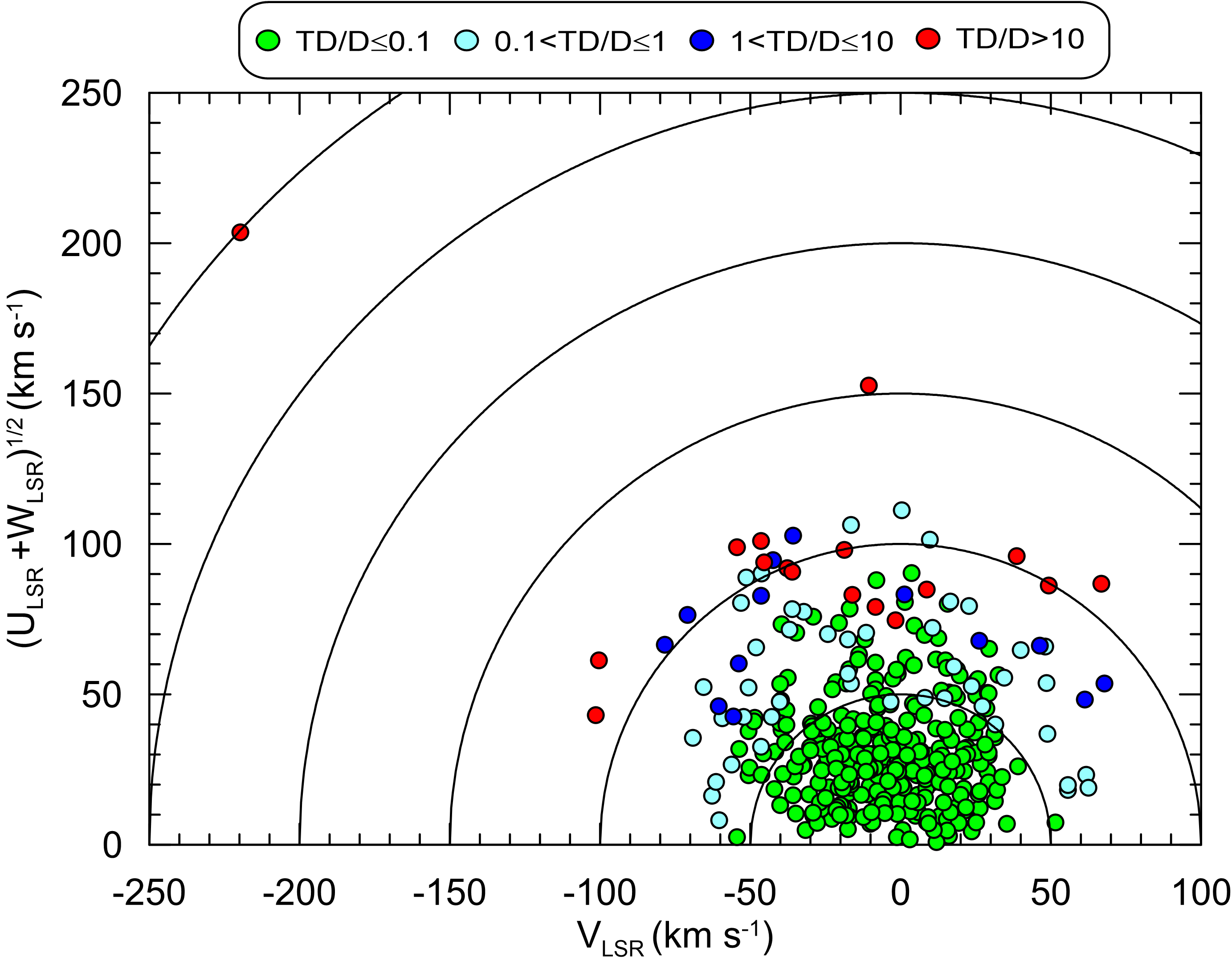}
   \caption{Toomre diagram of the final CV sample of 385 objects. Systems are colored according to their population types described in the text.}
   \label{fig:toomre}
\end{figure}

 Local densities of the Galactic components can be obtained from different objects in the solar neighborhood. For example, \citet{Chen2001} and \citet{Siegel2002} found 6.5–13\% and 6–10\%, respectively, for the local space density of the thick disk
 \citep[see also][and references therein]{Karaali2004, Bilir2006c, Bilir2008, Cabrera-Lavers2007}. \citet{Chen2001} and \citet{Siegel2002} used field stars from The Sloan Digital Sky Survey \citep[SDSS,][]{Yorketal2000} and Kapteyn selected areas, respectively. These values are in agreement with the thick disk percentage found in this study. Considering also the local thick-to-thin disk density normalization of $\rho_{\rm thick}/\rho_{\rm thin}=12\%$ and local halo-to-thin disk density normalization of 0.05\% given by \citet{Juric2008}, it is concluded that the systems in the sample are mostly members of the thin disk population, as expected from the spatial distribution of CVs in the sample (Figure \ref{fig:XYZ}). The total space velocity dispersions of the CV populations are listed in Table 3 together with their errors. The Toomre diagram of CVs in the final sample is shown in Figure \ref{fig:toomre}. In this diagram, systems are coloured according to their population types. 

\subsection{Orbital Periods}

\begin{figure}
  \centering
  \includegraphics[width=\columnwidth]{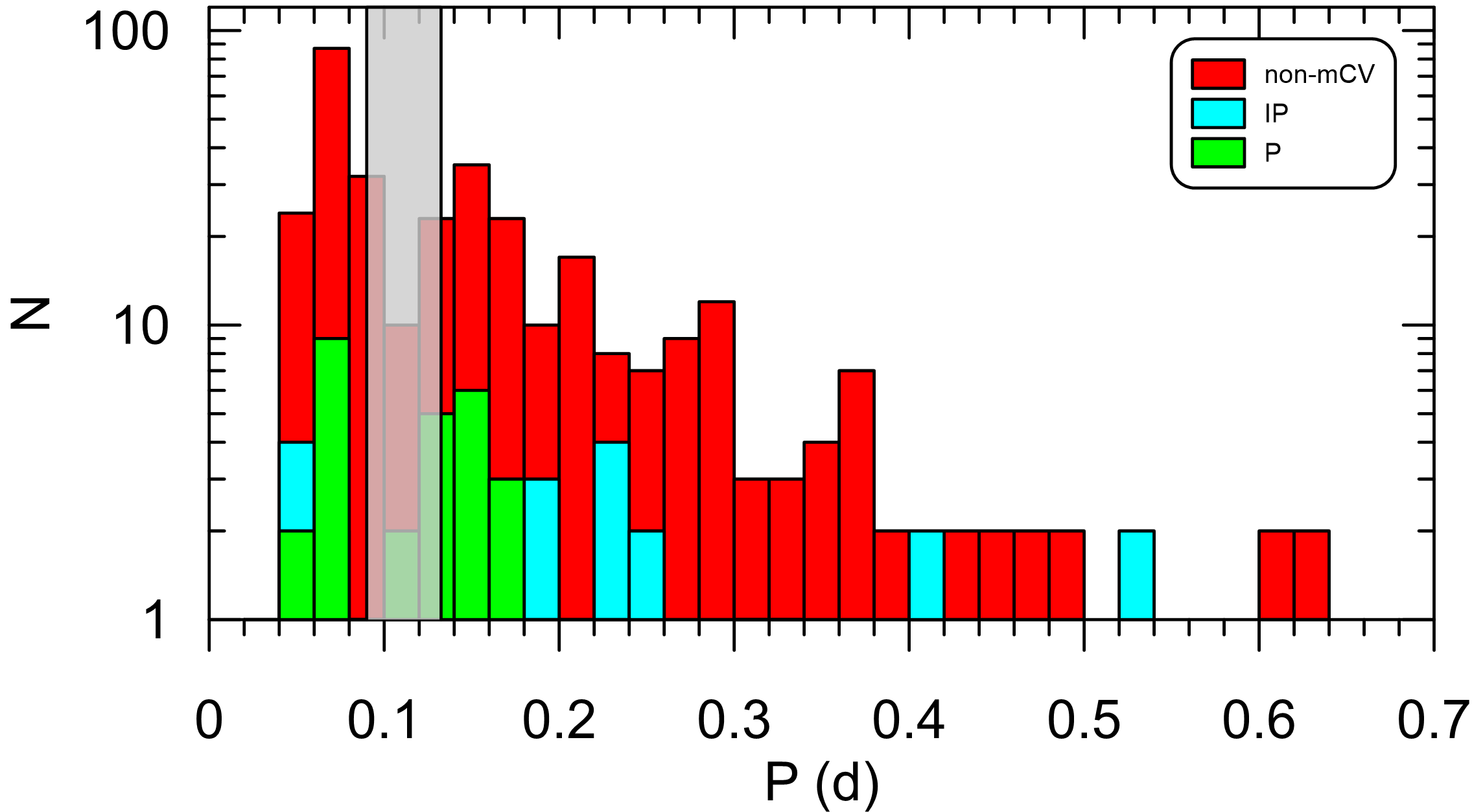}
   \caption{Histogram of orbital periods of the final CV sample including 385 CVs. The shaded area represents the orbital period gap. The vertical axis is given on a logarithmic scale since the numbers of polars (P) and intermediate polars (IP) are small compared to non-magnetic systems.}
   \label{fig:period_gap}
\end{figure}

\begin{figure}
  \centering
  \includegraphics[width=\columnwidth]{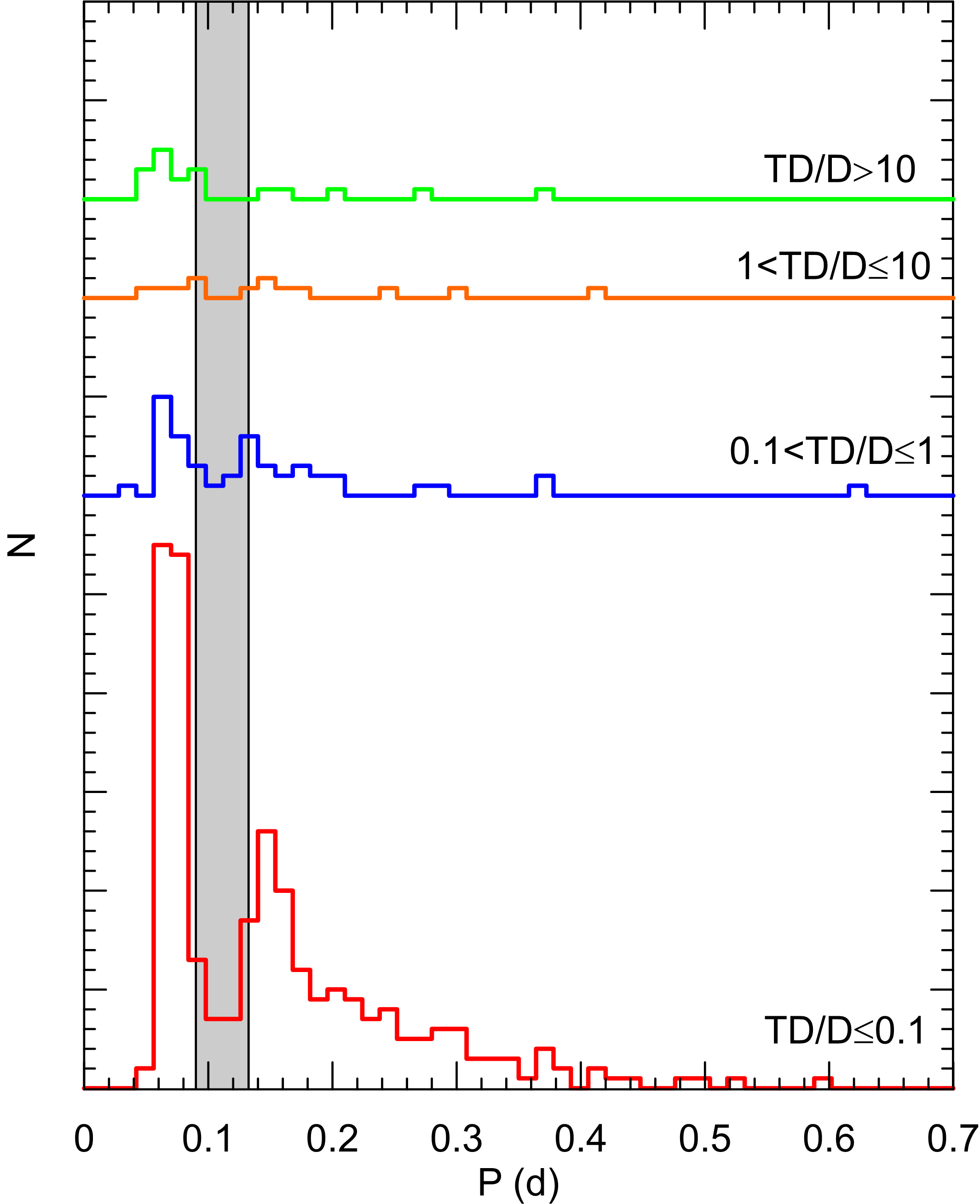}
   \caption{Histogram of orbital periods of the final CV sample. Systems are coloured according to their population types described in the text.}
   \label{fig:TD_his}
\end{figure}

An orbital period is the most precisely determined parameter for a CV. The orbital period histogram of 385 CVs in the final sample is presented in Figure \ref{fig:period_gap}, where the orbital period gap between 2.15 and 3.18 hr \citep{Knigge2006, Knigge2011b} makes the period histogram bimodal.  Ps, IPs, and non-mCVs are shown separately in Figure \ref{fig:period_gap}. Orbital period histograms plotted according to population types are shown in Figure \ref{fig:TD_his}, where the period gap is very prominent for CV groups with $TD/D\leq 0.1$ and $0.1<TD/D\leq 1$, high and low-probability thin disk stars. Hence, as they constitute a homogeneous sample, we decided to study the kinematical properties of high and low-probability thin disk  CVs only, for which the total number of systems is 354. 

Space velocity dispersions of all CVs and non-magnetic systems belonging to the thin disk component of the Galaxy are given in Table 3 for below and above the orbital period gap. Total space velocity dispersions of the thin disk CVs below and above the gap are 47.44$\pm$4.19 and 43.73$\pm$4.47 km~s$^{-1}$, respectively, where errors are standard deviations of individual space velocity errors. When magnetic CVs are removed from the sample, total space velocity dispersions of non-magnetic CVs belonging to the thin disk component are found to be 47.67$\pm$3.94 and 44.43$\pm$4.33 km~s$^{-1}$ for the systems below and above the gap, respectively. The total space velocity dispersion of  all magnetic CVs in the thin disk component is estimated at 42.68$\pm$5.37 km~s$^{-1}$. The total space velocity dispersion of polars and IPs in the thin disk were calculated separately and found to be 42.76$\pm$5.77 and 41.70$\pm$4.91 km~s$^{-1}$, respectively.

The sample of thin disk CVs including 354 systems were also investigated for total space velocity dispersion differences between the various orbital period regimes. To perform this investigation, the sample was divided into smaller subsamples of almost a similar number of CVs at different period ranges. Since the number of systems is being smaller for periods longer than 0.28 d, the number of systems is also smaller for the period regimes above 0.28 d. The total space velocity dispersions, $\gamma$ velocity dispersions, and kinematical ages are given in Table 3 for all the thin disk CVs and non-magnetic thin disk CVs grouped into various orbital period intervals. 

\subsection{On the Selection Biases and Completeness of the Sample}

Photometric data sets used in testing CV evolution models are affected by selection biases. The primary bias comes from the brightness-related selection effects \citep{Pretorius2007}.  

Because it is easier to obtain spectra of brighter stars, our sample is skewed towards brighter systems. This results in our sample mostly containing long-period objects that are close to the Sun, in general, since long-period CVs are brighter than those with short orbital periods. Indeed, Table 3 shows that the number of systems with $P<2.15$ h and $P>3.18$ h are 139 and 188, respectively. It seems that there is no crucial difference between the numbers of faint and brighter systems in the sample. In addition, most systems are located within 1 kpc of the Sun (see Figure \ref{fig:XYZ}). Because there is a statistically significant number of systems across orbital periods and magnetic nature (Figure \ref{fig:period_gap}), we expect there are enough systems across all sub-groups for a meaningful kinematic analysis. 

\begin{figure}
  \centering
  \includegraphics[width=\columnwidth]{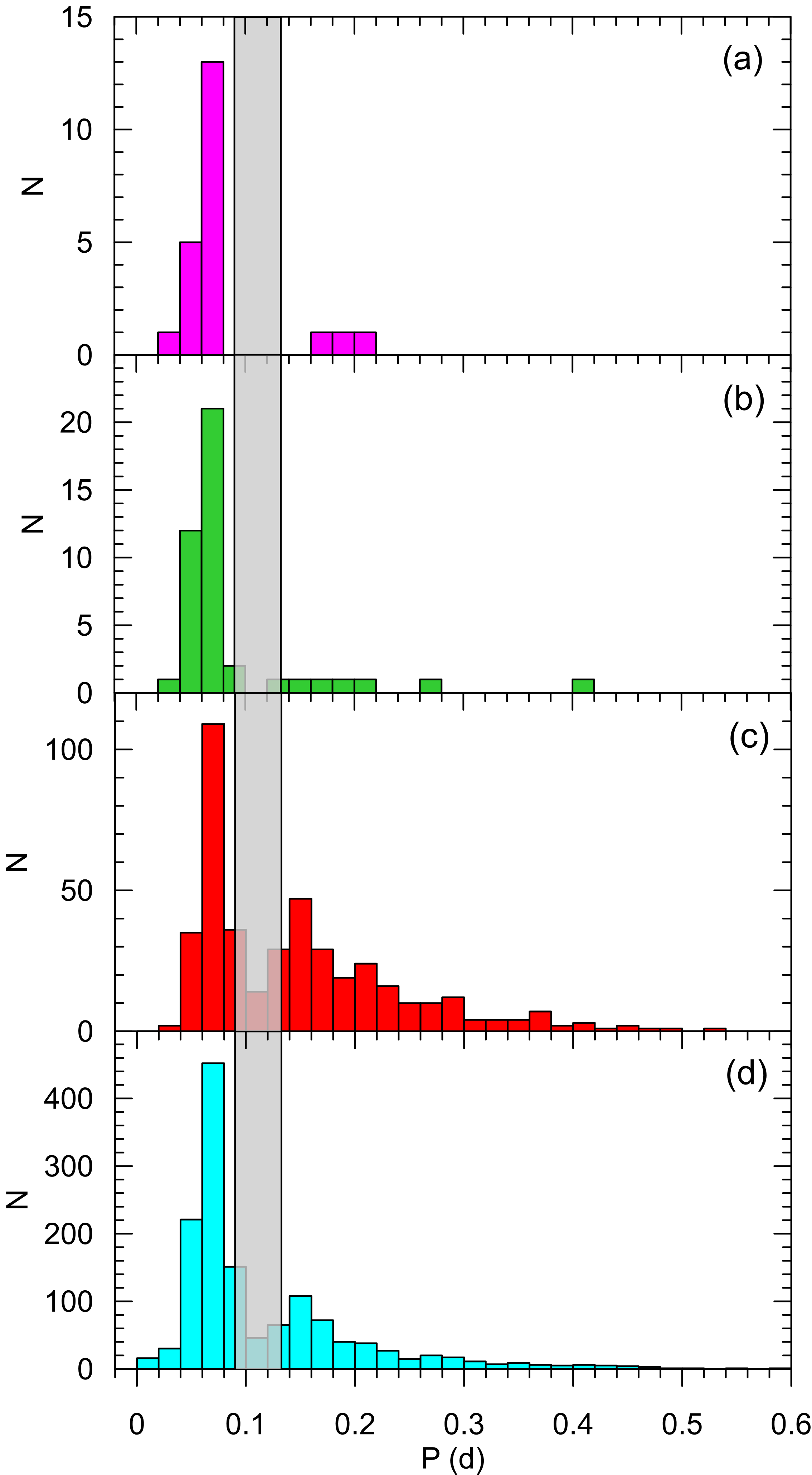}
   \caption{Comparison of orbital period histograms of the samples used in \citet{Rodriguezetal2024} (a), \citet{Pala2020} (b) and this study (c). Orbital period histogram of the catalogue of \citet{Ritter2003} is shown in the bottom panel. The shaded area represents the orbital period gap.}
      \label{fig:Per_H_P20}
\end{figure}

A comparison of the orbital period histograms of the samples used in \citet[][hereafter referred to as P20]{Pala2020}, \citep[][hereafter referred to as R24]{Rodriguezetal2024} and this study is shown in Figure \ref{fig:Per_H_P20}. The orbital period histogram of the catalogue of \citet{Ritter2003} is also shown in the figure. Note that R24 and P20 used volume-limited samples. Figure \ref{fig:Per_H_P20} reveals that the samples in P20 and R24 are biased towards the short-period (fainter) systems, while the sample in this study better represents the CVs with known orbital periods.  

Spatial distribution and Galactic positions of the objects can also produce a somewhat small bias in the final sample. This bias was tested using Figure \ref{fig:coordinates} and Figure \ref{fig:XYZ}, where positions of CVs are shown according to their equatorial and Galactic coordinates, and Sun-centred rectangular Galactic coordinates, respectively. Median values and standard deviations of the Sun-centred rectangular Galactic coordinates $X$, $Y$ and $Z$ are -82$\pm$503, 104$\pm$509 and 50$\pm$331 pc for all CVs in the final sample, respectively, while median value and standard deviation of Galactic latitudes are $4^{\circ}.72 \pm 31^{\circ}.68$. These values show that there is no considerable bias according to the spatial distribution of CVs in our study.

Observational studies of CVs suffer from strong selection biases not only due to brightness and spatial distribution but also due to different phenomenology. It can be found a detailed discussion of these biases in P20 (see also R24). When considering brightness-dependent selection biases, which is probably the strongest one, it is worth paying some more attention to the studies of P20, \citet[][hereafter referred to as C23]{Canbay2023} and R24, who estimated space densities of CVs from {\it Gaia} DR2, {\it Gaia} DR3 and X-ray surveys, respectively. P20 and R24 obtained space densities of 4.8$^{+0.6}_{-0.8}\times$10$^{-6}$ pc$^{-3}$ and 3.7$(\pm0.7)\times$10$^{-6}$ pc$^{-3}$ from their volume-limited CV samples, respectively, which are in good agreement. Their samples include objects within the sphere with a radius of 150 pc with the Sun at the centre. P20 point out that their volume-limited sample is 77$\pm$10\% complete and it must contain 12 more CVs to obtain a 100\% complete sample. They give their final space density estimation as 4.8$^{+0.6}_{-0.8}\times$10$^{-6}$ pc$^{-3}$ by using this assumed 100\% complete sample of CVs. 

On the contrary, C23 did not use a volume-limited sample. Instead, they limited their sample by defining the completeness limits of the data for certain absolute magnitude intervals. The median value of CV distances in their sample is 989 pc, with distances extending to 5-6 kpc from the Sun. This median value is about 6.5 times larger than the upper distance limit of the volume-limited samples mentioned here. The space density of CVs estimated by C23 is 6.8$^{+1.3}_{-1.1}\times$10$^{-6}$ pc$^{-3}$, which is surprisingly in good agreement with those given by P20 and R24. This value is even higher than the estimates given in the aforementioned studies. It should be noted that P20 and R24 assumed a scale height of $H$ = 280 pc in their studies, while C23 simultaneously estimated the space density and scale height ($H$ = 375 pc). 

Since the main purpose of this work is to study the general kinematic properties of CVs rather than focus on 100\% completeness, the larger sample of C23 is most appropriate for this work. 

White dwarf kicks can be considered to add some bias to the space velocity dispersions. Simulations performed by \citet{El-Badry2018} show that a modest velocity kick with typical velocities of $\sim$0.75 km s$^{-1}$ is possible during white dwarf formation, probably due to asymmetric mass-loss. Adding this small velocity to the total space velocity dispersion will affect the age. For example, the final sample of 385 CVs in Table 3 has a total space velocity dispersion of 54.29$\pm$4.51 km s$^{-1}$, corresponding to a mean kinematical age of 5.40$\pm$0.83 Gyr. An increase of 0.75 km s$^{-1}$ in this total space velocity dispersion has a very small effect on the total dispersion and age calculations, resulting in $54.04\pm 4.52$ km s$^{-1}$ and 5.53$\pm$0.83 Gyr. We conclude that the possible bias from the white dwarf kicks can be ignored, unless substantial evidence for larger WD kicks during CV formation are presented in the future.

\begin{figure}
  \centering
  \includegraphics[width=\columnwidth]{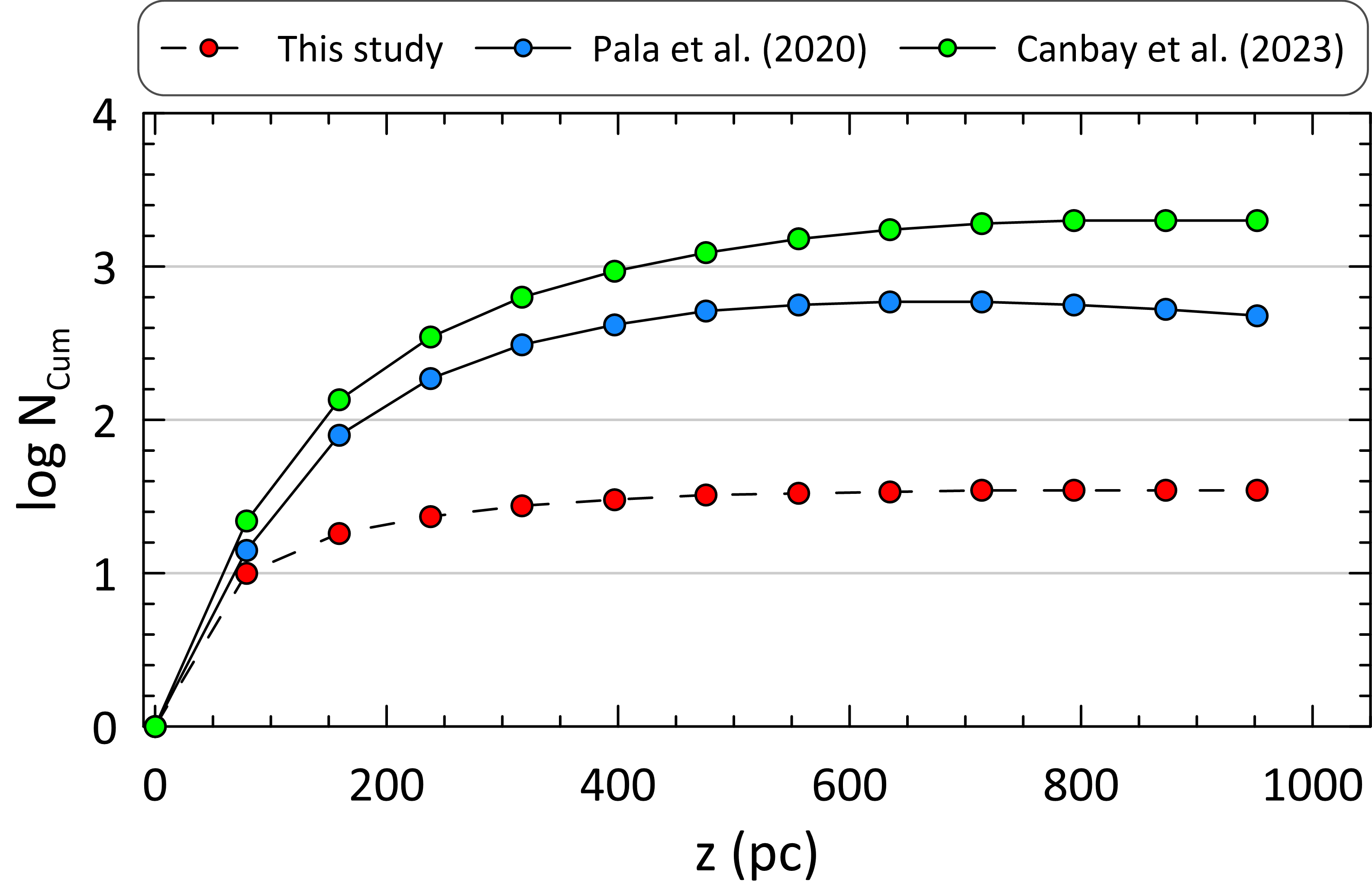}
   \caption{Comparison of the cumulative number of CVs estimated for 100 pc intervals. The numbers were estimated using the space densities and scale heights given by \citet{Pala2020} and \citet{Canbay2023}. Red circles represent the data sample in this study.}
      \label{fig:Den_Comp}
\end{figure}

In order to test the completeness of our sample, we found the expected distributions of the cumulative number of CVs from the space density estimations given by P20 and C23, and compared them with the distribution of systems in our sample. Expected numbers of objects can be estimated by utilizing the exponential density equation of the thin disk population: 

\begin{equation}
D(x,z) = n \times \exp \left(-\frac{\mathopen|z+z_{0}\mathclose|}{H}\right)\times\exp \left(-\frac{(x-R_{0})}{h}\right)
\end{equation}
where $n$ is the normalized density of CVs in the solar neighborhood, $z$ and $z_0$ are the distances of the CV being studied and the Sun from the Galactic plane \citep[$z_0=27\pm 4$ pc,][]{Chen2000}, respectively, $H$ and $h$ scale height and scale length of CVs, $R_0$ solar distance from the Galactic centre \citep[$R_0$ = 8 kpc,][]{Majewski1993}. $x$ is the planar distance of the object from the Galaxy centre and calculated as, 

\begin{equation}
x = \left[R_{0}^{2} + (z/\tan b)^{2} - 2R_{0}(z/\tan b)\cos l\right]^{1/2}
\end{equation}
where $l$ and $b$ are the Galactic longitude and latitude of the object. Since the scale length of CVs is not known, we could estimate the number of CVs from the space densities for $l$=0$^{\circ}$ and $b$= 90$^{\circ}$, for the vertical distances from the Galactic plane. Thus, the exponential density equation takes the following form,

\begin{equation}
D(z) \cong n \times \exp \left(-\frac{\mathopen|z+z_{0}\mathclose|}{H}\right).
\end{equation}
We estimated the theoretical number of stars for 100 pc intervals up to 1200 pc distance. Since these distances were designed for $b=90^{\circ}$ and $z=d\times\sin b$, we get $z=d$. Centroids of distance intervals of 100 pc are defined as $d^{*}=z^{*}= (d_{1}^{3}+d_{2}^{3})/2$. In this study, we take $d_{1}$ = 0 pc. Then, we determined space densities for each selected distance using space densities in the solar neighborhood and scale heights given by P20 and C23. The volume element increases with increasing distance, thus, we used the equation $V_{i}=(4/3)\pi d_{i}^{3}$ to calculate the volume corresponding to the distance $d_{i}$. The expected number of CVs was calculated by multiplying the volume element and its related density. The estimated numbers of CVs for 100 pc intervals in the direction ($l$, $b$) = (0$^{\circ}$, 90$^{\circ}$) were found for 4.8$\times$10$^{-6}$ pc$^{-3}$ and $H=280$ pc from P20 and 6.8$\times$10$^{-6}$ pc$^{-3}$ and $H$=375 pc from C23. Cumulative numbers of CVs estimated in this way are plotted against distance from the Galactic plane ($z$) in Figure \ref{fig:Den_Comp}. For comparison, we used the CVs with $S_{\rm err}\leq 16.02$ km~s$^{-1}$ in this study. Since unit absolute magnitudes ($\Delta M_{G}$=1) must be considered in order to estimate the Galactic model parameters, the absolute magnitude interval of the sample must be determined. As our calculations show that $G$-magnitude changes between 0 and 11 mag, the effective unit absolute magnitude interval is 11 mag for the sample. Thus, CV numbers were normalized by dividing them by 11. The calculated cumulative number of CVs for our sample are shown in Figure \ref{fig:Den_Comp}. 

Figure \ref{fig:Den_Comp} reveals that our sample is not complete due to brightness-related selection effects, compared to the expected numbers of CVs from the space densities in P20 and C23. A statistical study based on this sample would be biased due to brightness-related selection effects. However, such effects can not create strong bias on our sample since we have sufficient systems across orbital periods and magnetic nature. All we need in a kinematical analysis is to have a sample including as many as objects possible, and precisely measured parameters.

\section{Discussions}

A sample of 432 CVs with systemic velocities and astrometric data was collected from literature and {\it Gaia} DR3 \citep{Gaia2023}, respectively. In order to minimize biases in the trigonometric parallax data of the {\it Gaia} DR3, we have used recalculated parallaxes in \citet{BailerJones2021} to obtain precise distances and their errors of the CV sample. We also limited the preliminary sample of 432 CVs with $S_{\rm err}$=16.02 km~s$^{-1}$, obtaining a final sample of 385 CVs.  There are 1,108 systems with distances calculated from {\it Gaia} DR3 database and \citet{BailerJones2021} parallaxes in the catalogue of \cite{Ritter2003}. By taking into account distance errors, the \cite{Ritter2003} catalogue includes 42 CVs within the sphere with a radius of 150 pc with the Sun at the centre, 182 CVs within 300 pc, 390 CVs within 500 pc and 810 CVs within 1 kpc. Our sample of 385 CVs contains 26 CVs (62\% of the \cite{Ritter2003} catalogue) within 150 pc, 97 CVs (53\%) within 300 pc, 179 CVs (46\%) within 500 pc and 332 CVs (41\%) within 1 kpc. Although our sample is not complete for a brightness-related study, kinematical studies can be very weakly affected by brightness-related biases \citep{Kolb2001} (see section 2.8). The median distance of the final sample is 570 pc. The present sample of this study occupies a much larger space than that in previous studies \citep{Van1996,Ak2010,Ak2015}. We believe that this data sample is the most reliable than ever used for kinematical analysis of CVs.

Objects in the final sample are located well within the Galactic disk in the solar neighborhood, as median values and standard deviations of their $X$, $Y$ and $Z$ coordinates are -82$\pm$503 pc, 104$\pm$509 pc and 50$\pm$331 pc, respectively (Figure~\ref{fig:XYZ}). The mean space velocities with respect to LSR of the final sample were found to be $\langle U_{\rm LSR} \rangle=1.08\pm3.65$ km s$^{-1}$, $\langle V_{\rm LSR}\rangle = -6.52\pm3.55$ km s$^{-1}$ and $\langle W_{\rm LSR}\rangle = -2.75\pm2.78$ km s$^{-1}$. Space velocity dispersions of this sample were calculated $\sigma_{\rm U_{LSR}} = 38.66\pm 2.80$ km s$^{-1}$, $\sigma_{\rm V_{LSR}} = 29.90\pm2.67$ km s$^{-1}$ and $\sigma_{\rm W_{ LSR}}=23.64\pm 2.31$ km s$^{-1}$ km s$^{-1}$. Total space velocity dispersion of all CVs in the final sample is $\sigma_{\nu} = 54.29\pm4.51$ km s$^{-1}$, indicating a $\gamma$ velocity dispersion of $\sigma_{\gamma}$ = 31.34$\pm$2.61 km s$^{-1}$ and a mean kinematical age of $\tau$=5.40$\pm$0.83 Gyr. These values and scatter in Figure \ref{fig:UVW_err} indicate that there are both younger and older systems in the final sample. Thus, in order to obtain homogeneous object groups and investigate their kinematical properties, it is necessary to form various sub-groups of the final sample.


\subsection{Groups According to Population Types}

The Toomre diagram in Figure \ref{fig:toomre} reveals that kinematical properties of stars are related to their Galactic populations, as the distribution of objects belonging to different Galactic population components shows very explicit differences. Thus, it is very important to determine the Galactic population types of stars in a kinematical analysis. We found that about \%92 of CVs in the final sample belong to the thin disk component of the Galaxy, while $\sim$\%8 of them belong to the thick disk. 

Local densities derived from different objects in the solar neighborhood have a wide range of values. \citet{Chen2001} and \citet{Siegel2002} found 6.5–13\% and 6–10\%, respectively, for the local space density of the thick disk \citep[see also][and references therein]{Karaali2004, Bilir2006c, Bilir2008, Cabrera-Lavers2007}. \citet{Juric2008} gives the local thick-to-thin disk density normalization $\rho_{\rm thick}/\rho_{\rm thin}=12\%$, while local halo-to-thin disk density normalization is 0.05\%. Thus, we concluded that the number density rates of thin and thick disk CVs in our study are in agreement with the above values. A similar result can be found in \citet{Ramsayetal2018} who used {\it Gaia} DR2 parallaxes and proper motions to determine the expected cumulative distribution of AM CVn stars (a subgroup of CVs including two white dwarfs in the binary system) and compare it with the distribution of a Galactic disk population. In their Table B.1, the numbers of thin disk, thick disk and thin/thick objects are 25, 11 and 4, respectively, corresponding to the percentages of 62.5\%, 27.5\% and 10\%. As can be seen that these percentages are not in agreement with those found from the different objects in the solar neighborhood. Although they found that the most of the systems in their sample are members of the thin disk component of the Galaxy, \citet{Ramsayetal2018} concluded that a significant number of AM CVn stars are likely awaiting to be discovered. We conclude that our sample better represents the population profile of the solar neighbourhood.

The total space velocity dispersions of the CVs in $TD/D\leq 0.1$, $0.1<TD/D\leq 1$, $1<TD/D \leq 10$ and $TD/D>10$ ranges were calculated as 40.21$\pm$4.38, 71.68$\pm$4.83, 82.94$\pm$5.06 and 124.71$\pm$4.36 km~s$^{-1}$, respectively. The $\gamma$ velocity dispersions corresponding to these groups are 23.22$\pm$2.52, 41.38$\pm$2.79, 47.89$\pm$2.92 and 72.00$\pm$2.52 km~s$^{-1}$, respectively, while kinematical ages are 2.89$\pm$0.72, 8.51$\pm$0.83, 10.37$\pm$0.80, and 16.03$\pm$0.50 Gyr, respectively. It is clear that CV groups in Table 3 according to population types exhibit an increase in dispersions and ages from high probability thin disk stars to high probability thick disk stars, as expected. 

A wide range of ages for the thin and thick disk components of the Galaxy is found in various studies. \citet{Liu2000} found the ages of the thin and thick disks to be 9.7$\pm$0.6 Gyr and $t\sim$12.5$\pm$1.5 Gyr, respectively, while \citet{delPeloso2005} and \citet{Grundahl2008} estimated 8.8$\pm$1.7 and 8-9 Gyr for age of the thin disk. In a recent study, \citet{Kilic2017} derived the ages of 6.8–7.0 Gyr and 8.7$\pm$0.1 Gyr for local thin and thick disks from the white dwarf luminosity functions, respectively. Kinematical ages estimated for CV populations are not in accord with those found in previous studies based on field stars and clusters. High probability thin disk CVs were found much younger than local thin disk. Since white dwarfs in CVs are more massive than single white dwarfs \citep[see][and references therein]{Zorotovic2020}, CVs could be younger than single white dwarfs due to faster evolution. However, high probability thick disk CVs were found much older than local thick disk and this idea does not work for high probability thick disk CVs, probably due to their number in our sample being only 13 and the results for thick disk CVs are unreliable. 

Population analysis and Figure \ref{fig:XYZ} show that almost all of the CVs in the final sample are systems belonging to the thin-disk component of the Galaxy.  The number of high and low-probability thin disk CVs in the final sample is 354, which constructs a homogeneous sample for kinematical analyses. Thus, high and low-probability thin disk CVs were also considered for further analyses.   

\subsection{Kinematics of Magnetic and Non-magnetic CVs}

Derived kinematical parameters are listed in Table 3 for magnetic and non-magnetic CVs in the thin disk component of the Galaxy. Since the number of mCVs is only 49, they were not divided into sub-groups.  There are only 4 polars and 1 intermediate polar belonging to the thick disk component of the Galaxy in the sample with $S_{\rm err} < 16.02$ km~s$^{-1}$.

The total space velocity dispersion of all thin disk mCVs is 42.68$\pm$5.37 km s$^{-1}$, corresponding to a mean kinematical age of 3.31$\pm$0.92 Gyr. This parameter was calculated for non-mCVs as 46.33$\pm$4.23 km~s$^{-1}$, with a mean kinematical age of 3.95$\pm$0.75 Gyr. It should be noted that \citet{Ak2015} calculated these parameters for mCVs and non-mCVs in the thin disk component of the Galaxy, as done in this study. They estimated total space velocity dispersions of mCVs and non-mCVs as 55.60$\pm$7.63 and 44.90$\pm$6.97 km~s$^{-1}$, respectively, with mean kinematical ages of 5.64$\pm$1.39 and 3.69$\pm$1.22 Gyr. A comparison shows that differences in dispersions and ages between mCVs and non-mCVs in this study are not as large as found in \citet{Ak2010,Ak2015}. Moreover, dispersions of mCVs are larger than that of non-mCVs in \citet{Ak2010,Ak2015}, contrary to this study. It was expressed that evolutionary scenarios of magnetic CVs may be different from scenarios of non-magnetic CVs \citep[see the references in][]{Belloni2020a}. Thus, a considerable difference is expected between total space velocity dispersions of mCVs and non-mCVs. However, the data sample in this study does not show a large difference between total space velocity dispersions, correspondingly ages, of mCVs and non-mCVs. Polars and intermediate polars represent different phases of CV evolution. Thus, we also calculated the total space velocity dispersions of polars and intermediate polars in the thin disk as 42.76$\pm$5.77 and 41.70$\pm$4.91 km~s$^{-1}$, respectively, with the ages of 3.32$\pm$0.99 and 3.14$\pm$0.83 Gyr. The kinematic properties of these magnetic CV groups belonging to the thin disk component of the Galaxy do not show a significant difference, although we believe that their numbers in our sample are still small.

\subsection{Groups According to Orbital Periods}

An orbital period is the most precisely determined parameter of a CV, and it is related to the mass of the secondary star \citep{Kalomeni2016}. As the secondary mass and orbital period are age-related quantities, kinematical parameters derived for the orbital period ranges could be meaningful. Since the orbital period evolution of CV samples has been predicted by the standard theory of evolution of these systems, any kinematical analysis based on orbital period ranges could be used to test evolutionary scenarios. 

In order to investigate kinematical differences between subgroups of orbital period ranges, systems were first divided into two subgroups according to the orbital period gap. Total space velocity dispersions of the thin disk CVs below and above the gap are 47.44$\pm$4.19 and 43.73$\pm$4.47 km~s$^{-1}$, respectively, corresponding to kinematical ages of 4.15$\pm$0.75 and 3.49$\pm$0.78 Gyr. When magnetic CVs are removed from the sample, total space velocity dispersions of non-magnetic CVs belonging to the thin disk component are 47.67$\pm$3.94 and 44.43$\pm$4.33 km~s$^{-1}$ for the systems below and above the gap, respectively, corresponding to kinematical ages of 4.19$\pm$0.71 and 3.61$\pm$0.74 Gyr. 

According to population synthesis models of Galactic CVs based on the standard formation and evolution theory, the mean age of systems below the gap is 3-4 Gyr, while the CVs above the period gap are younger than 1.5 Gyr \citep{Kolb1996, Ritter1986}. It is concluded that this age difference is mainly due to the time spent evolving from the post-CE phase into the contact phase \citet{Kolb2001}. The mean kinematical age calculated in this study for the systems below the gap is in agreement with the upper limit of the theoretical prediction, while the mean kinematical age for the CVs above the gap is not in agreement with the prediction. Note that the difference between the mean ages of systems below and above the period gap is not as large as expected from the population models of Galactic CVs based on the standard formation and evolution theory.  Our results show that the dispersion of CVs below the gap is larger than that of those above the gap. Note that this is at most a 2$\sigma$ result and the measured quantity is the total space velocity dispersion, not the age. Thus, expected and observational $\gamma$ velocity dispersions of systems below and above the gap should be also compared. 

CVs in the sample were grouped according to orbital period ranges to investigate the kinematic properties of these groups. The total space velocity dispersions and corresponding kinematical ages of CVs in these period ranges are listed in Table 3. The continuous decrease of the dispersion and age from shorter to longer orbital periods is remarkable. The results emphasize the importance of detecting stellar populations in kinematic analyses for CV samples. Since the sample including only non-magnetic thin disk CVs is supposed to be the most homogeneous sample, these objects were also grouped according to orbital period ranges. The total space velocity dispersions and corresponding kinematical ages of CVs in these period ranges are listed in Table 3. The same trend, decreasing total space velocity dispersions from shorter to longer periods, is clearly present in these data as well.   

Thus, we can conclude that the orbital period is decreasing with increasing age. The mean kinematical age increases from 2.83$\pm$0.73 to 4.29$\pm$0.66 Gyr, while the orbital period decreases from 0.550 to 0.030 d for the non-magnetic thin disk systems. This is not in agreement with the age-period relation in \citet{Ak2010}, but it is similar to the relation presented in \citet{Ak2015}. The age-orbital period relations found in this study are shown in Figure \ref{fig:Per_Age}a and b for all thin disk CVs and non-magnetic thin disk CVs, respectively, where red solid lines represent linear fits. Decreasing rates in Figure \ref{fig:Per_Age}a and b are $dP/dt=-1.82(\pm0.18)\times10^{-5}$ sec yr$^{-1}$ and $dP/dt=-2.09(\pm0.22)\times10^{-5}$ sec yr$^{-1}$, respectively. 

Decreasing orbital periods with increasing age is expected from the standard evolution model of CVs \citep[see][and references therein]{Kalomeni2016}. Assuming that the observed period changes in \citet{Schaefer2024} are due to evolutionary processes, an inspection of CVs with decreasing orbital period in Table 1 of \citet{Schaefer2024} shows that mean period decrease of WD-MS systems with orbital periods $0.0567 \leq P~(d) \leq 0.618$ is $dP/dt=-4.13(\pm0.44)\times10^{-6}$ sec yr$^{-1}$. Note that this period range is consistent with the sample in this study. The period change rate in our study is about 5 times higher than {that of the orbital periods} in \citet{Schaefer2024}.  However, it is not clear if the observed orbital period changes in \citet{Schaefer2024} originated from evolutionary processes, since \cite{Kingetal2024} claims that the observed period changes must be originated from short–term phenomena and long-term evolution of CVs can be constrained only by the observations with timescales longer than $\sim 3\times10^{5}$ yr.

\begin{figure}
  \centering
  \includegraphics[width=\columnwidth]{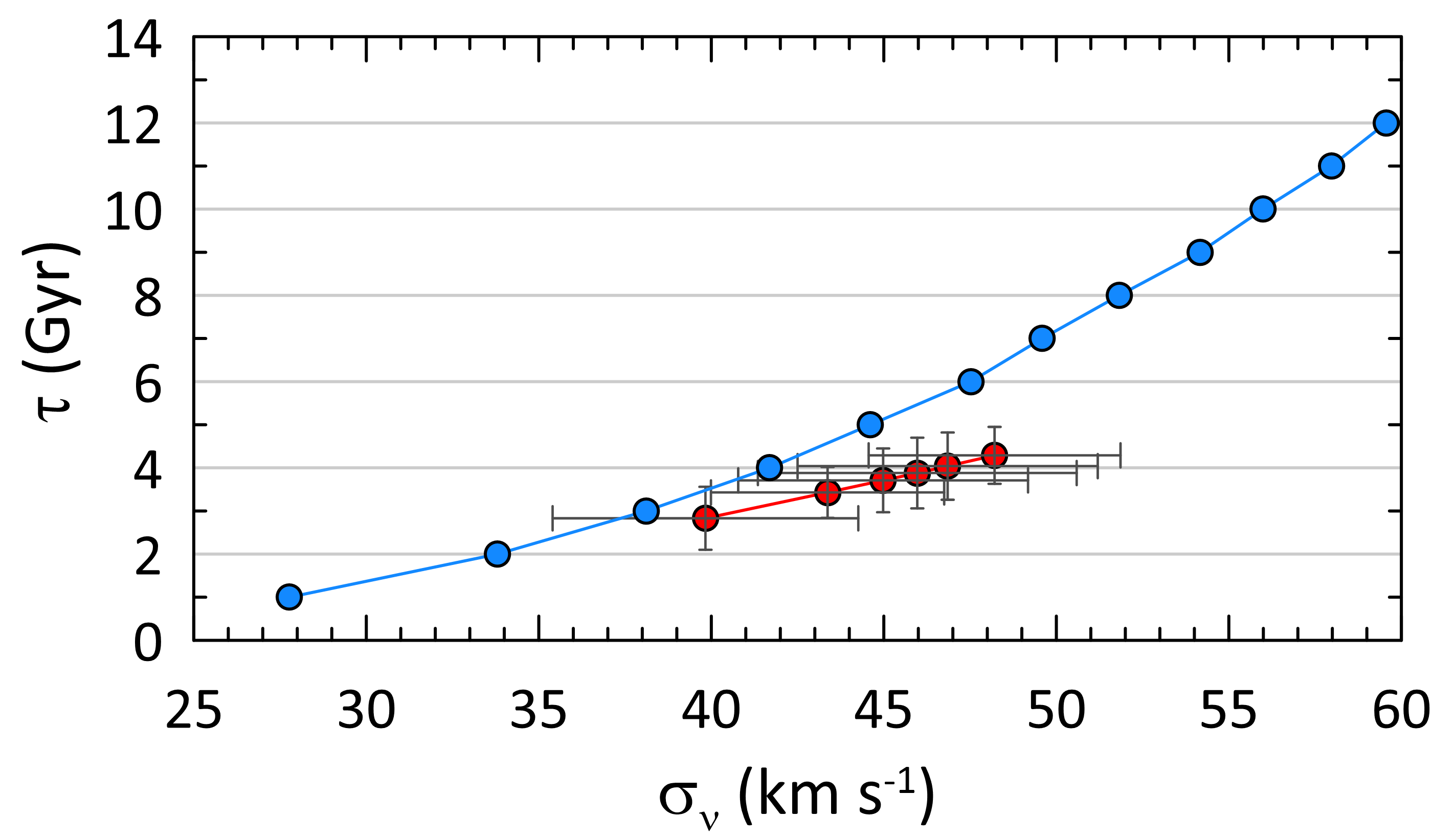}
   \caption{Comparison of age-velocity dispersion relations of CVs and single white dwarfs. Blue circles represent the relation obtained from \citet{Chengetal19}, while red circles show the values found for thin-disk CVs in this study.}
      \label{fig:WD_CV-comp}
\end{figure}

\begin{figure}
  \centering
  \includegraphics[width=\columnwidth]{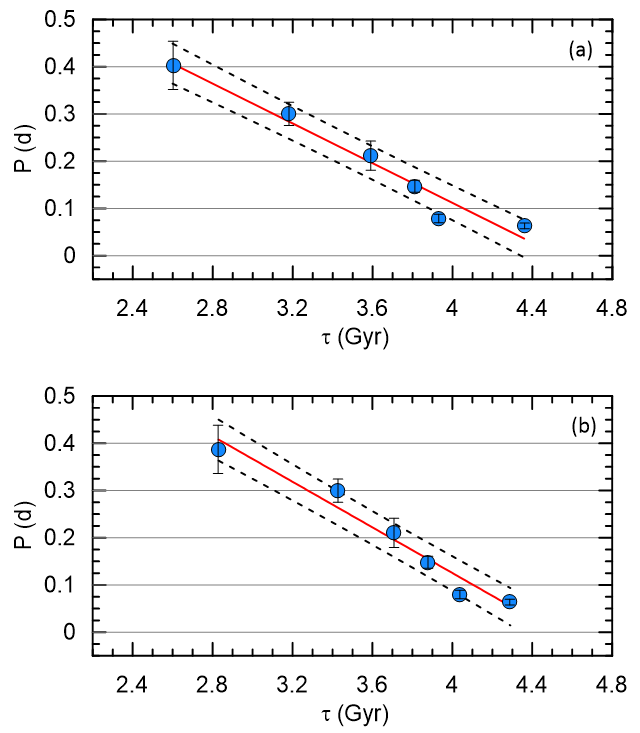}
   \caption{Orbital period-kinematic age relations found for all CVs (a) and non-magnetic CVs (b) located in the thin-disk component of the Galaxy. Red solid lines represent the linear fits.}
      \label{fig:Per_Age}
\end{figure}

 Age-velocity dispersion relation of CVs in the thin-disk component of the Galaxy can be compared with that of isolated white dwarfs given by \citet[][hereafter referred to as C19]{Chengetal19} (see also \citet{Raddietal22}) in Figure \ref{fig:WD_CV-comp}. For comparison, we have taken the $\sigma_{\rm U}$, $\sigma_{\rm V}$, $\sigma_{\rm W}$ and age values of isolated white dwarfs in C19. \citet{Raddietal22} compared their dispersion-age relation with that of C19 in their Figure 11, in which the relations in C19 are extrapolated in order to remove the thick disk contribution for ages higher than $\sim$7 Gyr. We used these extrapolated relations to obtain the total space velocity values and ages in C19. Using the extrapolated relations, we calculated total space velocity dispersions ($\sigma_{\nu}$) of single white dwarfs. Figure \ref{fig:WD_CV-comp} shows that the age-dispersion relation of CVs is not in agreement with that of isolated white dwarfs, although the youngest CVs (long-period systems) are partly in agreement. It seems that CVs are younger than isolated white dwarfs. However, considering that the age-dispersion relation in C19 includes both the thin and thick disk CVs, while the relation in this study represents only thin-disk systems, disagreement between the two relations should be less than in Figure \ref{fig:WD_CV-comp}.

\newpage

\subsection{The $\gamma$ Velocity Dispersions}

The $\gamma$ velocity dispersions of CV sub-groups can be used to test predictions of the standard formation and evolution theory of CVs. The $\gamma$ velocity dispersions of mCVs and non-mCVs belonging to the thin disk component of the Galaxy are very similar, $\sigma_{\gamma} = 24.64\pm3.10$ km s$^{-1}$ and $\sigma_{\gamma} = 26.75\pm2.44$ km s$^{-1}$, respectively.  Note that this is a 1$\sigma$ result.

When it is calculated only for thin disk non-mCVs, $\gamma$ velocity dispersions of the systems below and above the orbital period gap are found $\sigma_{\gamma} = 27.52\pm2.28$ km s$^{-1}$ and $\sigma_{\gamma} = 25.65\pm2.44$ km s$^{-1}$, respectively (Table 3). Dispersions of the $\gamma$ velocities are expected to be $\sigma(\gamma)\simeq 30$ km s$^{-1}$ and $\sigma(\gamma)\simeq 15$  for the systems below and above the period gap, respectively, according to the standard formation and evolution theory based on disrupted magnetic braking \citep{Kolb1996}. Observations can not reveal such a large difference between the $\gamma$ velocity dispersions of the systems located below and above the gap. 

It is clear that the results of the analysis in this study do not meet the expectations of the theory with respect to the $\gamma$ velocity dispersions. Also, previous observational kinematical studies do not satisfy the theory with respect to gamma velocities. On the other hand, \citet{Kolb2001} predicts that the $\gamma$ velocity dispersions are $\sigma(\gamma)\simeq 27$ and $\sigma(\gamma)\simeq 32$ km s$^{-1}$ for the systems above and below the gap, respectively, if magnetic braking does not operate in the detached phase during the evolution. Although observational results in Table 3 do not fully match this prediction,  while there is just a 3$\sigma$ disagreement, it is clear that the difference of the dispersions is smaller, as expected in Kolb's study (\citeyear{Kolb2001}). Similar results were obtained in previous studies \citep{Ak2010,Ak2015}. Another possible explanation of the present result may come from nova outbursts. We know that a longer life for a CV means more dynamical interaction with massive objects and irregular components of the Galactic potential. Thus, an older CV group must have a higher $\sigma_{\gamma}$ than younger object groups. However, if an object has undergone repeatedly to nova explosions with possibly asymmetric envelope ejection, then its space velocity may additionally be affected by the cumulative effect of these nova explosions. \citet{Kolb1996} states that even such a single intrinsic evolutionary event affects quadratically the dispersion of $\gamma$ velocity. As a result, $\sigma_{\gamma}$ values of older and younger CV groups can be very similar. Not to mention, the increase in the $\gamma$ velocity dispersions of CVs with the decreasing orbital period is clear (Table 3).  Although this trend is expected from the theory, calculated values do not verify the ones predicted by \citet{Kolb2001} for the CVs above the period gap, while there is only a 3$\sigma$ disagreement for the CVs below the gap.

\section{Conclusions}
With its most precise distances and proper motions data from {\it Gaia} DR3, and systemic velocities collected from the literature, the most reliable ever data sample of cataclysmic variables was constructed for their kinematical analysis. The sample is comprised of systems well within the Galactic disk in the Solar neighborhood. 

Total space velocity dispersion of all CVs in the final sample is $\sigma_{\nu} = 54.29\pm4.51$ km s$^{-1}$, indicating a $\gamma$ velocity dispersion of $\sigma_{\gamma}$ = 31.34$\pm$2.61 km s$^{-1}$ and a mean kinematical age of $\tau$=5.40$\pm$0.83 Gyr. 

 High probability thin disk CVs were found to be much younger than the local thin disk. Age of the local thin disk was derived 6.8–7.0 Gyr from the white dwarf luminosity functions \citep{Kilic2017}. CVs could be younger than single white dwarfs due to faster evolution, as CV white dwarfs are more massive compared to single white dwarfs.  

We have also concluded that there is no considerable difference between total space velocity dispersions of magnetic and non-magnetic CVs.  In addition, a meaningful difference could not be found between polars and IPs in the thin disk when comparing total space velocity dispersions and kinematical ages.

Total space velocity dispersions of non-magnetic CVs belonging to the thin disk component of the Galaxy were found to be 47.67$\pm$3.94 and 44.43$\pm$4.33 km~s$^{-1}$ for the systems below and above the period gap, respectively, corresponding to kinematical ages of 4.19$\pm$0.71 and 3.61$\pm$0.74 Gyr. The mean kinematical age calculated for the systems below the gap is in agreement with the upper limit of the theoretical prediction, while there is no agreement with the theoretical age prediction for the CVs above the gap. In addition, the difference between the mean ages of systems below and above the period gap could not be found as large as expected from the population models of Galactic CVs. 

It is also found in this study that the orbital period is decreasing with increasing age. The age-orbital period relations were found to be $dP/dt=-1.82(\pm0.18)\times10^{-5}$ s yr$^{-1}$ for all thin disk CVs, and $dP/dt=-2.09(\pm0.22)\times10^{-5}$ s yr$^{-1}$ for non-magnetic thin disk CVs. Such a trend is expected from the standard evolution model of CVs.

 A comparison of age-velocity dispersion relations of CVs in the thin disk component of the Galaxy with that of isolated white dwarfs given by \citet[][see also \citet{Raddietal22}]{Chengetal19} shows that CVs are younger than isolated white dwarfs. As presented here, this difference between isolated WDs and CVs is just over a 1$\sigma$ result, but after considering the previous point that would bring this difference closer, it is likely that there would be no significant difference in age between isolated WDs and CVs.

$\gamma$ velocity dispersions of the non-magnetic thin disk CVs below and above the orbital period gap were obtained $\sigma_{\gamma} = 27.52\pm2.28$ km s$^{-1}$ and $\sigma_{\gamma} = 25.65\pm2.44$ km s$^{-1}$, respectively. We could not find a large difference between the $\gamma$ velocity dispersions of the systems below and above the gap, contrary to the prediction of the standard formation and evolution theory \citet{Kolb1996}. This small difference between the dispersions is in agreement with \citet{Kolb2001}, in which it is expressed that the dispersion difference is small if magnetic braking does not operate in the detached phase during the evolution.  Note that the kick imparted by nova outbursts could be responsible for this result, as well.

\section*{Acknowledgments}
The authors thank the anonymous referee for his/her helpful comments that improved the quality of the manuscript. This study has partly been supported by the Scientific and Technological Research Council (TÜBİTAK) 119F072. This work has been supported in part by Istanbul University: project number NAP-33768. This study is a part of the PhD thesis of Remziye Canbay. This research made use of NASA's (National Aeronautics and Space Administration) Astrophysics Data System and the SIMBAD Astronomical Database, operated at CDS, Strasbourg, France and the NASA/IPAC Infrared Science Archive, which is operated by the Jet Propulsion Laboratory, California Institute of Technology, under contract with the National Aeronautics and Space Administration. This study used data from the European Space Agency (ESA) mission {\it Gaia} (\mbox{https://www.cosmos.esa.int/gaia}) and, processed by the {\it Gaia} Data Processing and Analysis Consortium (DPAC, \mbox{https://www.cosmos.esa.int/web/gaia/dpac/consortium}). Funding for the DPAC was provided by national institutions, particularly institutions participating in the {\it Gaia} Multilateral Agreement.


\bibliography{Canbay}
\bibliographystyle{aasjournal}

\end{document}